\documentclass[10pt]{article}
\usepackage[latin9]{inputenc}
\usepackage[T1]{fontenc}
\usepackage[french,english]{babel}
\usepackage{graphicx}
\usepackage[top=3cm, bottom=3cm, left=3cm, right=3cm]{geometry}
\usepackage{float}
\usepackage{amssymb}
\usepackage{amsmath}
\usepackage{titlepic}
\usepackage{caption} 
\usepackage{epstopdf}
\usepackage{xfrac}
\usepackage{enumitem}
\usepackage{subfigure}
\usepackage{bm}
\usepackage{esint}
\usepackage{mathtools} 
\usepackage{url}
\usepackage[dvipsnames]{xcolor}

\usepackage{amsthm}

\newcommand{\R}{\mathbb{R}}
\newcommand{\N}{\mathbb{N}}

\newcommand{\C}{\mathbb{C}}

\newcommand{\Dbf}{\mathbf{D}}

\newcommand{\dd}{\mathrm{d}}
\newcommand{\tin}{\text{ in }}

\newtheorem{theorem}{Theorem}[section]
\newtheorem{remark}{Remark}[section]
\newtheorem{corollary}{Corollary}[section]
\newtheorem{definition}{Definition}[section]
\newtheorem{lemma}{Lemma}[section]

\newtheorem{proposition}{Proposition}[section]

\newtheorem{cond}{Condition}

\begin{document}
	
	\title{Modal approximation for plasmonic resonators in the time domain: the scalar case}
	\author{Lorenzo Baldassari\thanks{\footnotesize Department of Mathematics, 
			ETH Z\"urich, 
			R\"amistrasse 101, CH-8092 Z\"urich, Switzerland.}  \and Pierre Millien\thanks{\footnotesize  Institut Langevin, 1 Rue Jussieu, 75005 Paris, France. }\and Alice L. Vanel\footnotemark[1] ${}^,$\thanks{Corresponding author: alice.vanel@sam.math.ethz.ch}}
	
	\date{\today}
	
	\maketitle
	
\begin{abstract}
We study the electromagnetic field scattered by a metallic nanoparticle with dispersive material parameters in a resonant regime. We consider the particle placed in a homogeneous medium in a low-frequency regime. We define modes for the non-Hermitian problem as perturbations of electrostatic modes, and obtain a modal approximation of the scattered field in the frequency domain. The poles of the expansion correspond to the eigenvalues of a singular boundary integral operator and are shown to lie in a bounded region near the origin of the lower-half complex plane. Finally, we show that this modal representation gives a very good approximation of the field in the time domain. We present numerical simulations in two dimensions to corroborate our results.
\end{abstract}
	
	\def\keywords2{\vspace{.5em}{\textbf{  Mathematics Subject Classification
				(MSC2000).}~\,\relax}}
	\def\endkeywords2{\par}
	\keywords2{35R30, 35C20.}
	
	\def\keywords{\vspace{.5em}{\textbf{ Keywords.}~\,\relax}}
	\def\endkeywords{\par}
	\keywords{plasmonic resonance, time-domain modal expansion, subwavelength resonators, quasi-normal modes}

\section{Introduction}

	\subsection{Context}
When describing the interaction of light with a resonating particle, summing the natural resonant modes of the system is an intuitive and attractive approach. The modes are easily computed as they are eigenmode solutions to a source-free problem. They are intrinsic quantities of the system and give insights to understand the underlying physics. Once they are calculated, the response of the system to any given excitation can be computed at a low computational cost. 
A bounded, lossless system is Hermitian and admits a basis of orthonormal eigenmodes associated to real eigenvalues. But for a system that exhibits loss (by absorption or radiation), the classical spectral theorem cannot be used to diagonalise the non-Hermitian operator and the eigenvalues become complex \cite{lalanneReview, lalanne_modal_18, PhysRevA.4.1782}.

Several authors have obtained modal expansions for non-Hermitian systems \cite{PhysRevApplied.11.044018, Ge:14, Kokkotas1999, Leung:96, Muljarov_2010, Pick:17, PhysRevB.90.075108, PhysRevA.89.023829, zworski1999resonances}. Their use in nanophotonics is quite recent and is studied by many research groups  in the physics community (see the review paper \cite{lalanneReview} and references therein). Nevertheless, a number of theoretical and numerical issues arise \cite{Colom18}. Modes of non-Hermitian systems are not orthogonal, using classical inner products. In order to satisfy the outgoing boundary conditions, these generalised modes have complex frequencies with negative imaginary parts and, if they decay exponentially in time as $t \to \infty$, they grow far away from the resonating systems. This is known in the literature as Lamb's exponential catastrophe \cite{sirenko2007modeling}. Recently, frameworks for the computation and normalisation of these generalised modes have been established in different settings \cite{PhysRevA.90.013834, PhysRevA.92.053810, Stout_2017, PhysRevA.89.023829, PhysRevB.91.195422,stout2019eigenstate}.

\subsection{Scope of the paper}

In this paper we consider the scattering of a scalar wave by an obstacle with dispersive parameters (described by a \emph{Drude-Lorentz model}). This is a good model for the scattering of light by a dispersive obstacle in the \emph{transverse magnetic polarisation} (see \cite[remark 2.1]{moiola2019acoustic}).
We work in a \emph{low-frequency regime} corresponding to relevant physical applications, such as the scattering of light in the visible/infrared domain by a metallic nanoparticle whose characteristic size is a few tens of nanometers. 

The goal of this paper is to obtain an \emph{approximation}  of the \emph{low-frequency part} of the scattered field by a dispersive obstacle in the time domain as a \emph{finite sum} of modes oscillating at complex frequencies.

The tools used are singular boundary integral equations and elementary functional analysis. In this paper we do not deal with the high frequency part of the field that is usually studied with micro-local analysis tools. 

\subsection{Previous work on plasmonic resonances and layer potentials}
It has been shown in \cite{ammari2016surface, ammari2017mathematicalscalar, ammari2016plasmaxwell} that using boundary integral representation and layer potential analysis, one can define the resonant frequencies as solutions of a non-linear eigenvalue problem on the boundary of the particle.
 In a low-frequency regime, i.e. at frequencies corresponding to wavelengths that are orders of magnitude larger than the particle's size, asymptotic analysis techniques, as in \cite{ammari2017mathematicalscalar}, yield a hierarchy of boundary integral equations. The asymptotic small parameter is $\delta\omega c^{-1}$, where $\delta$ is the size of the particle, $\omega$ the frequency and $c$ the velocity. At leading order the well-known Neumann-Poincar\'e operator appears \cite{S1904}. Using the Plemelj symmetrisation principle and the spectral theory of compact self-adjoint operators, the latter can be diagonalised in the appropriate functional spaces \cite{khavinson2007poincare,nedelec2001acoustic}, which allows the scattered field to be decomposed in a basis of orthogonal modes in the static case \cite{ando2016analysis}.
The properties of the eigenvalues of the Neumann-Poincar\'e operator have been extensively studied in the literature, see the review paper \cite{ando2020spectral} and references therein. For a smooth enough boundary, say $C^{1,\alpha}$ for some $\alpha>0$, the operator is compact and its eigenvalues are real numbers converging to zero. The eigenvalues of the Neumann-Poincar\'e operator in the two- and three-dimensional cases are intrinsically different. In two dimensions, the spectrum is symmetric with respect to the origin (except for the eigenvalue 1/2), so there are as many positive eigenvalues as negative. The decay rate of the eigenvalues depends strongly on the regularity of the boundary. For an analytic boundary, the eigenvalues have an exponential decay rate \cite{ando2018}. In three dimensions, very few surfaces are known to have negative eigenvalues \cite{ji19}. For a strictly convex $C^\infty$ domain, there are infinitely many positive eigenvalues and a finite number of negatives ones \cite{ando2020}. The eigenvalues rate of decay is much slower than in two dimensions: $\lambda_j=\mathcal{O}(j^{-1/2})$ as $j\rightarrow\infty$ \cite{miyanishi2020eigenvalues} and zero is not in the essential spectrum \cite{ando2020}.

\subsection{Contributions and organisation of the paper}
We begin by describing the problem geometry and we formulate the governing equations in section~\ref{sec:problem}. We introduce the layer potential and boundary integral formulation and recall the modal decomposition of the static ($\omega=0$) solution. In section \ref{sec:modal_decomp}, we prove that in three dimensions, for a strictly convex particle, the modal expansion can be truncated due to the super-polynomial decay of the expansion's coefficients.  With a perturbation argument, we deduce from the static ($\omega = 0$) result a modal approximation in the dynamic case (for a small non-zero frequency). The perturbation analysis yields size and frequency dependent dynamic complex resonant frequencies. We show that all the resonant frequencies have a negative imaginary part and lie in a bounded region near the origin.  
Finally, in section~\ref{sec:time}, using only elementary complex analysis techniques, we give an approximation for the low-frequency part of the scattered field in the time domain as a finite sum of modes oscillating at complex resonant frequencies. We also show with a simple causality argument that the \emph{exponential catastrophe} is not problematic in practice. In section~\ref{sec:num} we implement this expansion in the two-dimensional setting and illustrate the validity of our approach with numerical simulations.

\section{Problem geometry and formulation}\label{sec:problem}
	
\subsection{Problem setting}
We are interested in the scattering problem of an incident wave illuminating a plasmonic nanoparticle in $\R^d$, $d=2,3$. The homogeneous medium is characterised by electric permittivity $\varepsilon_m$ and magnetic permeability $\mu_m$. Let $D$ be a smooth bounded domain in $\R^d$, of class $C^\infty$, characterised by electric permittivity $\varepsilon_c$. We assume the particle to be non-magnetic, i.e., $\mu_c = \mu_m$. Let $D=z+\delta B$ where $B$ is the reference domain and contains the origin, and $D$ is located at $z\in\R^d$ and has a characteristic size $\delta \ll 1$. We define the wavenumbers $k_c=\omega\sqrt{\varepsilon_c\mu_c}$ and $k_m=\omega\sqrt{\varepsilon_m\mu_m}$. Let $\varepsilon=\varepsilon_c \chi(D)+\varepsilon_m \chi(\R^d \setminus \bar{D})$, where $\chi$ denotes the characteristic function. We denote by $c_0$ the speed of light in vacuum $c_0=1/\sqrt{\varepsilon_0\mu_0}$ and by $c$ the speed of light in the medium $c=1/\sqrt{\varepsilon_m\mu_m}$.
	
Hereafter we use the Drude model \cite{ordal83} to express the electric permittivity of the particle: 
	\begin{equation}\label{eq:drude}
	\varepsilon_c(\omega)=\varepsilon_0\left(1-\frac{\omega_p^2}{\omega^2+i\omega\mathrm{T}^{-1}}\right),
	\end{equation}
	where the positive constants $\omega_p$ and $\mathrm{T}^{-1}$ are the plasma frequency and the collision frequency or damping factor, respectively.

\begin{cond}\label{cond:2d}
In two dimensions, we assume the domain $D$ to be an algebraic domain of class $\mathcal{Q}$, i.e. a quadrature domain. An algebraic domain is a domain enclosed by a real algebraic curve, namely the zero level set of a bivariate polynomial. A quadrature domain is the conformal image of the unit disc by a rational function.
\end{cond}

\begin{remark}
Algebraic domains are dense among all planar domains, so every smooth curve can be described as a sequence of algebraic curves \cite{putinar19}.
\end{remark}

\begin{cond}\label{cond:3d}
In three dimensions, we assume the domain $D$ to be strictly convex: for any two points in $D$, the line segment joining them is contained in $D \setminus \partial D$.
\end{cond}
Throughout the rest of the paper, $D$ is assumed to satisfy conditions \ref{cond:2d} or \ref{cond:3d}.
	
\subsection{Helmholtz equation for a subwavelength resonator}
Given an incident wave $u^\text{in}$ solution to the Helmholtz equation, the scattering problem in the frequency domain can be modelled by
	\begin{equation}\label{eq:helmholtz}
	\nabla \cdot \frac{1}{\varepsilon(x)} \nabla u(x)+\omega^2\mu_m u(x) = 0,  \qquad x \in \R^d,
	\end{equation}
	subject to the Sommerfeld radiation condition
	\begin{equation*}
	\left| \frac{\partial(u-u^\text{in})}{\partial |x|}-ik_m(u-u^\text{in})\right|=\mathcal{O}\left(|x|^{-(d+1)/2}\right), \qquad \mbox{as } |x|\rightarrow \infty,
	\end{equation*}
	uniformly in $x/|x|$, for $\Re{k_m}>0$. The transmission conditions are given by
	\begin{equation*}
	\begin{dcases}
	\left.  u(x)\right|_+ =  \left.u(x)\right|_-,  & x \in \partial D,\\
	\left.\frac{1}{\varepsilon_m}\frac{\partial u(x)}{\partial \nu}\right|_+ = \left.\frac{1}{\varepsilon_c} \frac{\partial u(x)}{\partial \nu}\right|_-,& x \in \partial D.
	\end{dcases}
	\end{equation*}
	Here, $\partial \cdot / \partial \nu$ denotes the normal derivative on $\partial D$, and the $+$ and $-$ subscripts indicate the limits from outside and inside $D$, respectively.
	
	\begin{definition}
		We denote the contrast $\lambda$ by
		\begin{equation*}
		\lambda(\omega)=\frac{\varepsilon_m+\varepsilon_c}{2(\varepsilon_m-\varepsilon_c)}.
		\end{equation*}
	\end{definition}
	
		\begin{definition}[Resonant frequency, mode]
	We say $\omega$ is a \emph{resonant frequency} if there is a non-trivial solution to equation \eqref{eq:helmholtz} with $u^\text{in}=0$. We call the solution a mode.
	A \emph{subwavelength resonance} occurs when a resonant frequency $\omega$ satisfies $\omega\delta c^{-1} <1$. 
	\end{definition}

	\subsection{Layer potential formulation}
	Let $H^{1/2}(\partial D)$ be the usual Sobolev space and let $H^{-1/2}(\partial D)$ be its dual space with respect to the duality pairing $\left\langle \cdot,\cdot\right\rangle_{\frac{1}{2},-\frac{1}{2}}$. The field $u$ can be represented using the single layer potentials $\mathcal{S}^{k_c}_D$ and $\mathcal{S}^{k_m}_D$, introduced in definition~\ref{de:layerpotential}, as follows:
	\begin{equation} \label{eq:scalar_solution}
	u(x)=\begin{dcases}
	\mathcal{S}^{k_c}_D[\Phi](x), & x\in D,\\
	u^\text{in}(x)+\mathcal{S}^{k_m}_D[\Psi](x), & x\in \mathbb{R}^d \setminus \overline{D},
	\end{dcases}
	\end{equation}
	where the pair $(\Phi, \Psi) \in H^{-\frac{1}{2}}(\partial D)\times H^{-\frac{1}{2}}(\partial D)$ is the unique solution to
	\begin{equation}\label{eq:integraleqsystem}
	\begin{dcases}
	\mathcal{S}^{k_m}_D[\Psi](x)-\mathcal{S}^{k_c}_D[\Phi](x)=F_1,& x\in\partial D,\\
	\frac{1}{\varepsilon_m}\left(\frac{1}{2}I+\mathcal{K}^{k_m,*}_D\right)[\Psi](x)+\frac{1}{\varepsilon_c}\left(\frac{1}{2}I-\mathcal{K}^{k_c,*}_D\right)[\Phi](x)=F_2, & x\in\partial D,
	\end{dcases}
	\end{equation}
	and
	\begin{equation*}
	F_1=-u^\text{in}(x), \qquad F_2=-\frac{1}{\varepsilon_m}\frac{\partial u^\text{in}(x)}{\partial \nu}, \qquad x \in \partial D,
	\end{equation*}
	where $\mathcal{K}^{k_m,*}_D$ is the Neumann-Poincar\'e operator defined in definition~\ref{de:layerpotential}. The trace relations for the single layer potential are given in lemma~\ref{lem:symmetrisation}.

	\subsection{Scaling and small-volume approximation}
The goal of this section is to establish an equivalent formulation for \eqref{eq:integraleqsystem} in the form $\mathcal{A}^{\frac{\omega\delta}{c}}[\Psi]= F$ (proposition \ref{prop:scalar_psi}), in order to write an asymptotic expansion of the operator $\mathcal{A}^{\frac{\omega\delta}{c}}$ (lemma \ref{le:small_vol_A}) and a spectral decomposition for the limiting operator $\mathcal{A}^{0}$ (proposition \ref{prop:spectralA0}).
	The scaling is new in this context, but the asymptotic expansion and the spectral decomposition were first obtained in \cite{ammari2017mathematicalscalar}. We recall them here for the sake of completeness. The proofs are quite lengthy and technical, so they are included in the appendix. 
	
	Recall that $z$ is the centre of the resonator and $\delta$ its radius. We introduce the scaling $x=z+\delta X$. For each function $\Xi$ defined on $\partial D$, we define a corresponding function on $\partial B$ by $\widetilde{\Xi}(X) := \Xi(z+\delta X)$, $X \in \partial B$. The scaling properties of the integral operators are given in appendix~\ref{sec:scaling}. The solution $\widetilde{u}$ becomes
	\begin{equation}
	\label{eq:single-layer}
	\widetilde{u}(X)=\begin{dcases}
	\delta\mathcal{S}^{k_c\delta}_B[\widetilde{\Phi}](X), & X\in B,\\
	u^\text{in}(z+\delta X)+\delta\mathcal{S}^{k_m\delta}_B[\widetilde{\Psi}](X), & X\in \R^d \setminus \overline{B},
	\end{dcases}
	\end{equation}
	where the single-layer potential $\mathcal{S}^{k\delta}_B$ and Neumann-Poincar\'e operator $\mathcal{K}^{k\delta,*}_B$ are defined by the fundamental solution $\Gamma^{k\delta}$. The density pair $(\widetilde{\Phi},\widetilde{\Psi}) \in H^{-\frac{1}{2}}(\partial B)\times H^{-\frac{1}{2}}(\partial B)$ is the unique solution to
	\begin{equation*}
	\begin{dcases}
	\mathcal{S}^{k_m\delta}_B[\widetilde{\Psi}](X)-\mathcal{S}^{k_c\delta}_B[\widetilde{\Phi}](X)=\frac{1}{\delta} \widetilde{F}_1, & X\in\partial B,\\
	\frac{1}{\varepsilon_m}\left(\frac{1}{2}I+\mathcal{K}^{k_m\delta,*}_B\right)[\widetilde{\Psi}](X)+\frac{1}{\varepsilon_c}\left(\frac{1}{2}I-\mathcal{K}^{k_c\delta,*}_B\right)[\widetilde{\Phi}](X)=\widetilde{F}_2,& X\in\partial B,
	\end{dcases}
	\end{equation*}
	and
	\begin{equation*}
	\widetilde{F}_1=-u^\text{in}(z+\delta X), \qquad \widetilde{F}_2=-\frac{1}{\delta\varepsilon_m}\frac{\partial u^\text{in}(z+\delta X)}{\partial \nu_X}, \qquad X \in \partial B.
	\end{equation*}
	Since $\mathcal{S}_{B}^{k_c\delta}:H^{-1/2} (\partial B)\rightarrow H^{1/2}(\partial B)$ is invertible for $k_c\delta$ small enough (see lemmas~\ref{lem:invertibility3D} and~\ref{lem:invertibility2D}), the following proposition holds.
	
	\begin{proposition} \label{prop:scalar_psi}
		For $d=2,3$, the following equation holds for $\widetilde{\Psi}$:
		\begin{equation}\label{eq:scalarpsi}
		\mathcal{A}_B^{\omega\delta/c}[\widetilde{\Psi}]=\widetilde{F},
		\end{equation}
		where
		\begin{eqnarray}
		\mathcal{A}_B^{\omega\delta/c}&=&\frac{1}{\varepsilon_m}\left(\frac{1}{2}I+\mathcal{K}^{k_m\delta,*}_B\right)+\frac{1}{\varepsilon_c}\left(\frac{1}{2}I-\mathcal{K}^{k_c\delta,*}_B\right)\left(\mathcal{S}^{k_c\delta}_B\right)^{-1}\mathcal{S}^{k_m\delta}_B, \notag \\ \label{eq:F_scalar}
		\widetilde{F}&=&\widetilde{F}_2+\frac{1}{\delta \varepsilon_c}\left(\frac{1}{2}I-\mathcal{K}^{k_c\delta,*}_B\right)\left(\mathcal{S}^{k_c\delta}_B\right)^{-1}[\widetilde{F}_1]. 
		\end{eqnarray}
	\end{proposition}

	\begin{lemma}[small-volume expansion]\label{le:small_vol_A}
		As $\omega\delta c^{-1}\rightarrow 0$, $\mathcal{A}_B^{\omega\delta/c}$ admits the following asymptotic expansion:
		\begin{equation}\label{eq:A_B}
		\mathcal{A}_B^{\omega\delta/c}=
		\begin{dcases}
		\mathcal{A}_B^{0}+\left(\omega\delta c^{-1}\right)^2\log{\left(\omega\delta c^{-1}\right)}\mathcal{A}_{B,1}+\mathcal{O}\left(\left(\omega\delta c^{-1}\right)^2\right), & d=2, \\
		\mathcal{A}_B^{0}+\left(\omega\delta c^{-1}\right)^2\mathcal{A}_{B,2}+\mathcal{O}\left(\left(\omega\delta c^{-1}\right)^3\right), & d=3,
		\end{dcases}
		\end{equation}
		where 
		\begin{equation}\label{eq:A_0}
		\mathcal{A}_B^{0}=\left(\frac{1}{2\varepsilon_c}+\frac{1}{2\varepsilon_m}\right)I-\left(\frac{1}{\varepsilon_c}-\frac{1}{\varepsilon_m}\right)\mathcal{K}^*_B,
		\end{equation}
		\begin{equation*}
		\mathcal{A}_{B,1} =\frac{1}{\varepsilon_m} \mathcal{K}^{(1)}_{B,1}(I -\mathcal{P}_{\mathcal{H}^*_0}) +  \left(\frac{1}{2}I - \mathcal{K}^*_B\right) \widetilde{\mathcal{S}}_B^{-1} \mathcal{S}_{B,1}^{(1)} \left(\frac{1}{\varepsilon_c}I - \frac{1}{\varepsilon_m } \mathcal{P}_{\mathcal{H}^*_0} \right),
		\end{equation*}
		and
		\begin{equation*}
		\mathcal{A}_{B,2}=\frac{\varepsilon_m-\varepsilon_c}{\varepsilon_m\varepsilon_c}\left(\frac{1}{2}I-\mathcal{K}^*_B\right)\mathcal{S}_B^{-1}\mathcal{S}_{B,2},
		\end{equation*}
		where the operators $\mathcal{P}_{\mathcal{H}^*_0}, \; \widetilde{\mathcal{S}}_B, \; \mathcal{S}_{B,1}^{(1)}, \;  \mathcal{S}_{B,2}$ and $\mathcal{K}^{(1)}_{B,1}$ are defined in appendix~\ref{subsec:exp_operators}.
	\end{lemma}
	
	\begin{proof}
		See appendix~\ref{subsec:A}.
	\end{proof}

The operator $\mathcal{A}_B^{\omega\delta/c}$ is not self-adjoint in $L^2$ so it can not be diagonalised directly to solve \eqref{eq:scalarpsi}. However, in the static regime, the operator  $\mathcal{A}_B^0$ can be expressed simply with $\mathcal{K}^*_B$, which can be symmetrised in the Hilbert space $\mathcal{H}^*(\partial B)$ (see appendix \ref{subsec:symm}).

\begin{lemma}[spectral decomposition of $\mathcal{K}_B^*$]\label{lem:spectralK}
	$\mathcal{K}^*_B$ is self-adjoint with respect to the inner product $\left\langle \cdot,\cdot\right\rangle_{\mathcal{H}^*(\partial B)}$. Moreover, it is compact, so its spectrum is discrete. The spectral theorem yields the decomposition
		\[
		\mathcal{K}^*_B=\sum_{j=0}^{+\infty} \lambda_j\left\langle \cdot,\widetilde{\phi}_j\right\rangle_{\mathcal{H}^*(\partial B)}\widetilde{\phi}_j,
		\]
		where $\{\lambda_j\}_{j\in\mathbb{N}}$ are the eigenvalues of $\mathcal{K}^*_B$ and $\{\widetilde{\phi}_j\}_{j\in\mathbb{N}}$ their associated normalised eigenvectors.
\end{lemma}

	\begin{proposition}[spectral decomposition of $\mathcal{A}_B^0$] \label{prop:spectralA0}The operator $\mathcal{A}_B^0$ has the spectral decomposition
		\begin{equation*}
		\mathcal{A}_B^0=\sum_{j=0}^{+\infty} \tau_j\left\langle \cdot,\widetilde{\phi}_j\right\rangle_{\mathcal{H}^*(\partial B)}\widetilde{\phi}_j,
		\end{equation*}
where $(\lambda_j,\widetilde{\phi}_j)_{j\in\N}$ are the eigenvalues and normalised eigenfunctions of $\mathcal{K}_B^*$ in $\mathcal{H}^*(\partial B)$ and
		\begin{equation*}
		\tau_j=\left(\frac{1}{\varepsilon_c}-\frac{1}{\varepsilon_m}\right)\left(\lambda(\omega)-\lambda_j\right).
		\end{equation*}
	\end{proposition}

	\begin{proof}
	Direct consequence of lemma \ref{lem:spectralK} and \eqref{eq:A_0}.
	\end{proof}
	
	\begin{corollary}
	The spectral approximation of the static ($\omega=0$) solution is given by
	\begin{equation*}
	\widetilde{u}(X)-\widetilde{u}^\text{in}(X)  =
	\sum_{j=0}^{\infty}\frac{1}{\tau_j} \left\langle \widetilde{F},\widetilde{\phi}_j\right\rangle_{\mathcal{H}^*(\partial B)} \delta \mathcal{S}_B[\widetilde{\phi}_j](X), \qquad X \in \mathbb{R}^d  \setminus \overline{B},
	\end{equation*}
	where $\widetilde{F}$ is defined in proposition \ref{prop:scalar_psi}.		
	
\end{corollary}

	\section{Modal decomposition of the field}
	\label{sec:modal_decomp}
	
In this section we want to apply perturbation theory tools to express the solutions of  \eqref{eq:scalarpsi} in terms of the eigenvectors of $\mathcal{K}^*_B$ that appear in the spectral decomposition of the limiting problem in proposition \ref{prop:spectralA0}, and to replace $\tau_j$ by a perturbed value $\tau_j(\omega\delta c^{-1})$.
Classical perturbation theory will give us a Taylor expansion for $\tau_j(\omega\delta c^{-1})$ in $\omega\delta c^{-1}$ for any $j\in \N$ but the remainders and validity range of these expansions will depend on the index $j$ of the considered eigenvalue. In order to get a meaningful expansion of the scattered field we need to work with a finite number of modes. 

\subsection{Modal expansion truncation} \label{subsec:truncate}
In practice, there is no need to consider the whole spectral decomposition of the field. It has been empirically reported that only a few modes actually contribute to the scattered field. The number of modes to consider increases as the source gets closer to the particle. In this section we give a mathematical explanation of this phenomenon : the modes $\widetilde{\phi}_j$ are eigenmodes of a pseudo-differential operator of order $-1$, and are oscillating functions. As in classical Fourier analysis, the decay with $j$ of the coefficients $\langle \widetilde{F},\widetilde{\phi}_j\rangle_{\mathcal{H}^*(\partial B)}$ will be determined by the regularity of the function $\widetilde{F}$ and the number of modes to consider will depend on the spatial variations of $\widetilde{F}$ over $\partial B$. In an homogeneous medium the incoming field is smooth and therefore we can expect a fast decay of the coefficients. 

\subsubsection{The three-dimensional case}

%\textcolor{red}{We can stay on $B$ and not worry about all the $\widetilde{\,}$ and just recall that $H^*$ is defined in appendix }
\begin{proposition} \label{prop:polyndecay}For $B$, a strictly convex domain in $\R^3$ with $C^\infty$-smooth boundary, and $\widetilde{F}\in H^J(\partial B)$ for some $J \in \N^*$ we have :\begin{equation}
	\left\langle \widetilde{F}, \widetilde{\phi}_j\right\rangle_{\mathcal{H}^*(\partial B)}=o(j^{-J/4}) \text{~as~} j\rightarrow +\infty.
	\end{equation}
\end{proposition}

The proof relies on a theorem from \cite{ando2020} which itself uses the computation of the principal symbol of the Neumann-Poincar\'e operator done in \cite{miyanishi2018weyl}:
\begin{theorem}[from \cite{ando2020}, p. 7]\label{theo:kang}
	For $B$, a strictly convex domain in $\R^3$ with $C^\infty$-smooth boundary, $\mathcal{K}_B^*$ has a finite number of non-positive eigenvalues. We can modify $\mathcal{K}_B^*$ by adding a finite dimensional smoothing operator to have a positive definite elliptic pseudo-differential operator of order -1, which we denote by $\widetilde{\mathcal{K}}_B^*$. For each real number $s\in\R$ there exist constants $c_s,C_s\in\R^+$ such that
	\begin{equation}\label{eq:ineq}
	c_s||\widetilde{\phi}||_{H^{s-1/2}(\partial B)} \leq ||\widetilde{\mathcal{K}}_B^*[\widetilde{\phi}]||_{H^{s+1/2}(\partial B)}\leq C_s||\widetilde{\phi}||_{H^{s-1/2}(\partial B)} 
	\end{equation}
	for all $\widetilde{\phi} \in H^{s-1/2}(\partial B)$. Moreover there exists $j_0\in \mathbb{N}$ such that
	\begin{equation*}
	\widetilde{\mathcal{K}}_B^*[\widetilde{\phi}_j]=\mathcal{K}_B^*[\widetilde{\phi}_j] \quad  \text{and}\quad  \lambda_j>0 \quad\text{for all }\quad  j \geq j_0.
	\end{equation*}
\end{theorem}

\begin{corollary}The operator $\mathbf{K}^*_B: L^2(\partial B) \longrightarrow L^2(\partial B)$ defined by $\mathbf{K}^*_B:=\left(-\mathcal{S}_B\right)^{\frac{1}{2}} \mathcal{K}_B^* \left(-\mathcal{S}_B\right)^{-\frac{1}{2}}$ is self-adjoint and has the same eigenvalues as $\mathcal{K}^*_B$. Its eigenvectors are $\widetilde{\psi}_j=\left(-\mathcal{S}_B\right)^{\frac{1}{2}}[\widetilde{\phi}_j]$. It can be modified by adding a finite dimensional smoothing operator to have a positive definite elliptic pseudo-differential operator of order -1, which we denote by $\widetilde{\mathbf{K}}_B^*$. For each real number $s\in\R$ there exist constants $c_s,C_s\in\R^+$ such that
	\begin{equation}\label{eq:ineq2}
	c_s||\widetilde{\phi}||_{H^{s-1/2}(\partial B)} \leq ||\widetilde{\mathbf{K}}_B^*[\widetilde{\phi}]||_{H^{s+1/2}(\partial B)}\leq C_s||\widetilde{\phi}||_{H^{s-1/2}(\partial B)} 
	\end{equation}
	for all $\widetilde{\phi} \in H^{s-1/2}(\partial B)$. Moreover there exists $j_0\in \mathbb{N}$ such that
	\begin{equation*}
	\widetilde{\mathbf{K}}_B^*[\widetilde{\psi}_j]=\mathbf{K}_B^*[\widetilde{\psi}_j] \quad  \text{and}\quad  \lambda_j>0 \quad\text{for all }\quad  j \geq j_0.
	\end{equation*}
\end{corollary}
\begin{proof}
	$\mathbf{K}^*_B$ has the same principal symbol as $\mathcal{K}_B^*$ \cite[p. 8]{miyanishi2020eigenvalues}. 
\end{proof}
We will also need the decay estimate of the eigenvalues of $\mathcal{K}_B^*$:
\begin{theorem}[from \cite{miyanishi2020eigenvalues}]\label{theo:asymptVP}
	For $B$, a strictly convex domain in $\R^3$ with $C^\infty$-smooth boundary the eigenvalues of the Neumann-Poincar\'e operator satisfy:
	\begin{align*}
	\lambda_j \sim C_B j^{-1/2},
	\end{align*} with $C_B$ a constant depending only on $B$: \begin{align*}
	C_B=\left(\frac{3 W(\partial B)-2\pi\chi(\partial B) }{128\pi}\right),
	\end{align*}
	where $W(\partial B)$  and $\chi(\partial B)$ denote, respectively, the Willmore energy and the Euler characteristic of the boundary surface $\partial B$.
\end{theorem}

\begin{proof}[Proof of proposition \ref{prop:polyndecay}]
	
	Consider $\widetilde{F}\in H^J(\partial B)$.  Since $\widetilde{\mathbf{K}}^*_B$ is a positive definite elliptic self-adjoint pseudo-differential operator of order $-1$ we can write \cite[p. 290]{demailly1997complex}:
	\begin{align*}
	H^{s}(\partial B) = \widetilde{\mathbf{K}}_B^*\left(H^{s-1}(\partial B)\right) \oplus \text{Ker\,}\left(\widetilde{\mathbf{K}}^*_B\right),
	\end{align*}
	where $\text{Ker\,}(\widetilde{\mathbf{K}}_B)$ denotes the kernel of $\widetilde{\mathbf{K}}_B^*$. The symbol $\oplus$ is to be understood in the $L^2$ scalar product sense. Hence for $j\geq j_0$:
	\begin{align*}
	\left\langle \widetilde{F}, \widetilde{\phi}_j\right\rangle_{\mathcal{H}^*(\partial B)} =& - \left\langle \widetilde{F}, \mathcal{S}_B[ \widetilde{\phi}_j]\right\rangle_{L^2(\partial B)} \\
	=&- \left\langle \widetilde{F}, \left(-\mathcal{S}_B\right)^{\frac{1}{2}}[ \widetilde{\psi}_j]\right\rangle_{L^2(\partial B)} \\
	=&- \left\langle \left(-\mathcal{S}_B\right)^{\frac{1}{2}}[ \widetilde{F}],\widetilde{\psi}_j\right\rangle_{L^2(\partial B)}.
	\end{align*} where we used the fact that $\left(-\mathcal{S}_B\right)^{\frac{1}{2}}$ is self-adjoint in $L^2(\partial B)$.
	Since $ \left(-\mathcal{S}_B\right)^{\frac{1}{2}}[ \widetilde{F}] \in H^{J+\frac{1}{2}}(\partial B)$ we have $ \left(-\mathcal{S}_B\right)^{\frac{1}{2}}[ \widetilde{F}]=\widetilde{\mathbf{K}}_B^*[\widetilde{G}^{(1)}] + \widetilde{F}^{(1)}_{\mathrm{ker}}$ with $\widetilde{G}^{(1)}\in H^{J-\frac{1}{2}}(\partial B)$. Then
	\begin{align*}
	\left\langle \left(-\mathcal{S}_B\right)^{\frac{1}{2}}[ \widetilde{F}],\widetilde{\psi}_j\right\rangle_{L^2(\partial B)}=& \left\langle\widetilde{\mathbf{K}}_B^*[\widetilde{G}^{(1)}] + \widetilde{F}_{\mathrm{ker}}^{(1)} , \widetilde{\psi}_j\right\rangle_{L^2(\partial B)} \\
	=& \lambda_j \left\langle \widetilde{G}^{(1)}, \widetilde{\psi}_j \right\rangle_{L^2(\partial B)} + \left\langle \widetilde{F}^{(1)}_{\mathrm{ker}} ,  \widetilde{\psi}_j\right\rangle_{L^2(\partial B)}.
	\end{align*}
	Since the eigenvectors of $\widetilde{\mathbf{K}}_B^*$ are orthogonal in $L^2(\partial B)$ we have:
	\begin{align*}
	\left\langle \widetilde{F}, \widetilde{\phi}_j\right\rangle_{\mathcal{H}^*(\partial B)} =- \lambda_j \left\langle \widetilde{G}^{(1)} ,\widetilde{\psi}_j\right\rangle_{L^2(\partial B)}.
	\end{align*}
	We can now write $\widetilde{G}^{(1)}= \widetilde{\mathbf{K}}_B^*[\widetilde{G}^{(2)}] + \widetilde{F}^{(2)}_{\mathrm{ker}}$ with $\widetilde{G}^{(2)}\in H^{J-\frac{3}{2}}(\partial B)$ and we have 
	\begin{align*}
	\left\langle \widetilde{G}^{(1)} ,\widetilde{\psi}_j\right\rangle_{L^2(\partial B)} = \lambda_j \left\langle \widetilde{G}^{(2)} ,\widetilde{\psi}_j\right\rangle_{L^2(\partial B)}.
	\end{align*}
	Iterating this procedure $J-1$ times yields
	\begin{align*}
	\left\langle \widetilde{G}^{(1)} ,\widetilde{\psi}_j\right\rangle_{L^2(\partial B)} = \lambda_j^{J-1} \left\langle \widetilde{G}^{(J)} ,\widetilde{\psi}_j\right\rangle_{L^2(\partial B)}.
	\end{align*}
	Hence\begin{align}\label{eq:iterateimage}
	\left\langle \widetilde{F}, \widetilde{\phi}_j\right\rangle_{\mathcal{H}^*(\partial B)} = -\lambda_j^J \left\langle \widetilde{G}^{(J)} ,\widetilde{\psi}_j\right\rangle_{L^2(\partial B)}.
	\end{align}
	We need to control the $L^2$-norm of $\widetilde{G}^{(J)}$.
	We can rewrite the orthogonal decomposition as  $\left(-\mathcal{S}_B\right)^{\frac{1}{2}}[\widetilde{F}] = \left(\widetilde{\mathbf{K}}_B^*\right)^J[\widetilde{G}^{(J)}]+ \widetilde{F}_{\mathrm{ker}}^{(1)}$. Composing by $\widetilde{\mathbf{K}}_B^*$ we get:
	\begin{align*}
	\widetilde{\mathbf{K}}_B^*\circ \left(-\mathcal{S}_B\right)^{\frac{1}{2}}[ \widetilde{F}] =  \left(\widetilde{\mathbf{K}}_B^*\right)^{J+1}[\widetilde{G}^{(J)}].
	\end{align*}
	Using the right-hand side of \eqref{eq:ineq2} with $s=J+\frac{1}{2}$ we get
	\begin{align*}
	\left\Vert  \left(\widetilde{\mathbf{K}}_B^*\right)^{J+1}[\widetilde{G}^{(J)}] \right\Vert_{H^{J+1}(\partial B)} \leq C_{J+\frac{1}{2}} \left\Vert  \left(-\mathcal{S}_B\right)^{\frac{1}{2}}[ \widetilde{F}]\right\Vert_{H^J(\partial B)}.
	\end{align*}
	%hence \begin{align*}
	%\left\Vert \left(\widetilde{\mathbf{K}}_B^*\right)^J[G^{(J)}]\right\Vert_{H^{J}} \leq \left\Vert \mathcal{S}^{\frac{1}{2}}_B[ F]\right\Vert_{H^{J}}.
	%\end{align*}
	Using $J+1$ times the left hand side of \eqref{eq:ineq2}  with $s-\frac{1}{2}=0,1,\ldots,J$ yields
	\begin{align*}
	\left\Vert \widetilde{G}^{(J)}\right\Vert_{L^2(\partial B)} \leq & \left(\prod_{s=0}^{J} \frac{1}{c_{s+\frac{1}{2}}} \right)\left\Vert \left(\widetilde{\mathbf{K}}_B^*\right)^{J+1}[\widetilde{G}^{(J)}]\right\Vert_{H^{J+1}(\partial B)} \\
	\leq & C_{J+\frac{1}{2}}\left(\prod_{s=0}^{J} \frac{1}{c_{s+\frac{1}{2}}} \right) \left\Vert \left(-\mathcal{S}_B\right)^{\frac{1}{2}}[ \widetilde{F}]\right\Vert_{H^{J}(\partial B)}.
	\end{align*}
	Using the Cauchy-Schwartz inequality in \eqref{eq:iterateimage}  and the fact that $\Vert \widetilde{\psi}_j\Vert_{L^2(\partial B)}=1$ ($\mathcal{S}_B$ is an isometry):
	\begin{align*}
	\left\vert \left\langle \widetilde{F}, \widetilde{\phi}_j\right\rangle_{\mathcal{H}^*(\partial B)} \right\vert \leq C\lambda_j^J \Vert \widetilde{F} \Vert_{H^{J-\frac{1}{2}}(\partial B)},
	\end{align*}
	where $C=C_{J+\frac{1}{2}}\left(\prod_{s=0}^{J} \frac{1}{c_{s+\frac{1}{2}}} \right)$ is independent of $j$.
	Using theorem \ref{theo:asymptVP} we can see that for $j$ large enough since $\lambda_j \sim C_B j^{-1/2}$ we have:
	\begin{align*}
	\left\vert \left\langle \widetilde{F}, \widetilde{\phi}_j\right\rangle_{\mathcal{H}^*(\partial B)} \right\vert \leq  j^{-J/2}C (C_B)^J \Vert \widetilde{F} \Vert_{H^{J-\frac{1}{2}}(\partial B)},
	\end{align*} and since $j^{-J/2} C (C_B)^J = o\left(j^{-J/4}\right)$ we get the result.
\end{proof}

\subsubsection{The two-dimensional case}
In two dimensions, the picture is a slightly different. Indeed, zero is in the essential spectrum of $\mathcal{K}_D^*$. The eigenspace associated to zero has infinite dimension and there are infinitely many negative eigenvalues. As a result, $\mathcal{K}_D^*$ can not be modified into a positive operator by a finite dimensional operator. However, for a certain class of domains, it is possible to show that there is a finite number of plasmonic resonances. For example, it was shown in \cite{putinar19} that an algebraic domain of class $\mathcal{Q}$ has asymptotically a finite number of plasmonic resonances. The asymptotic parameter is the deformation from the unit circle. For a larger class of domains the decay of the coefficients $ \langle F, \widetilde{\phi}_j\rangle_{\mathcal{H}^*(\partial D)} $ can be checked numerically (see section \ref{sec:num}).

\subsection{Modal decomposition}

Since the incoming wave is solution of the homogeneous Helmholtz equation in the background medium, standard elliptic regularity theory gives us $u^\text{in}\in C^\infty (\R^d)$. Moreover, the particle $B$ is assumed to be $C^\infty$, so the source term in equation \eqref{eq:scalarpsi}, i.e. the function $\widetilde{F}$, is smooth on $\partial B$. Therefore using proposition \ref{prop:polyndecay} we have a super-polynomial decay of the coefficients $\langle\widetilde{F}, \widetilde{\phi}_j\rangle_{\mathcal{H}^*(\partial B)}$, and we can consider that only a finite number of modes are excited. The number $J$ of modes to consider depends on the incoming field.

	\begin{proposition}
	Assume that $\widetilde{F}= \sum_{j=1}^J \left\langle\widetilde{F}, \widetilde{\phi}_j\right\rangle_{\mathcal{H}^*(\partial B)} \widetilde{\phi}_j$ on $\partial B$ for some $J\in \N^*$.
		The spectral approximation of the scattered field as $\omega\delta c^{-1}\rightarrow 0$ is given by
		\begin{equation*}
		\widetilde{u}(X)-\widetilde{u}^\text{in}(X)  =
		\sum_{j=0}^{J}\frac{1}{\tau_j(\omega)} \left\langle \widetilde{F},\widetilde{\phi}_j\right\rangle_{\mathcal{H}^*(\partial B)} \delta \mathcal{S}^{k_m\delta}_B[\widetilde{\phi}_j](X), \qquad X \in \mathbb{R}^d  \setminus \overline{B},
		\end{equation*}
		where
		\begin{equation*}
		\tau_j(\omega)=
		\begin{dcases}
		\tau_j+\left(\omega \delta c^{-1}\right)^2 \log{\left(\omega \delta c^{-1}\right)}\tau_{j,1}+\mathcal{O}\left(\left(\omega \delta c^{-1}\right)^2\right), & d=2, \\
		\tau_j+\left(\omega \delta c^{-1}\right)^2\tau_{j,2}+\mathcal{O}\left(\left(\omega \delta c^{-1}\right)^3\right), & d=3,
		\end{dcases}
		\end{equation*}
		with
		\[
		\tau_{j,1}= \left\langle \mathcal{A}_{B,1}\widetilde{\phi}_j,\widetilde{\phi}_j\right\rangle_{\mathcal{H}^*(\partial B)}, \qquad 
		\tau_{j,2}= \left\langle \mathcal{A}_{B,2}\widetilde{\phi}_j,\widetilde{\phi}_j\right\rangle_{\mathcal{H}^*(\partial B)},
		\] 
and $\widetilde{F}$ is defined in proposition \ref{prop:scalar_psi}.		
		
	\end{proposition}
	
	\begin{proof}
		Note that $\{\widetilde{\phi}_j\}_{j\in\mathbb{N}}$ forms an orthonormal basis of $\mathcal{H}^*(\partial B)$. Writing $\left(\mathcal{A}_B^0+\mathcal{A}_B^{\omega\delta/c}-\mathcal{A}_B^0\right)[\widetilde{\Psi}]=\widetilde{F}$ and using the decomposition of $\widetilde{\Psi}$ in $\mathcal{H}^*(\partial B)$, $\widetilde{\Psi}=\sum_{j=0}^{+\infty} \left\langle \widetilde{\Psi},\widetilde{\phi}_j\right\rangle_{\mathcal{H}^*(\partial B)}\widetilde{\phi}_j$, yields the following:
		\begin{equation*}
		\left\langle\widetilde{\Psi},\widetilde{\phi}_j\right\rangle_{\mathcal{H}^*(\partial B)}=
\left\{ \begin{aligned} & \frac{1}{\tau_j+\left\langle\left(\mathcal{A}_B^{\omega\delta/c} - \mathcal{A}_B^{0} \right)\widetilde{\phi}_j,\widetilde{\phi}_j\right\rangle_{\mathcal{H}^*(\partial B)}}\left\langle \widetilde{F},\widetilde{\phi}_j\right\rangle_{\mathcal{H}^*(\partial B)} \qquad & j\leq J \\
&0 \qquad & j>J.
\end{aligned} \right.
		\end{equation*}
		Using \eqref{eq:scalar_solution} and \eqref{eq:A_B} concludes the proof.
	\end{proof}

%	\begin{remark} \label{re:modes}
%		Note that $v_j=\mathcal{S}^{k_m\delta}_B[\widetilde{\phi}_j](X)$ solves the following
%		\begin{equation}
%		\begin{dcases}
%		\Delta v_j+(k_m\delta)^2 v_j=0, &\mbox \qquad \text{in}~ B~\text{and in}~\R^d \setminus \overline{B},\\
%		\left.\frac{\partial v_j}{\partial \nu}\right|_+ -\left.\frac{\partial v_j}{\partial \nu}\right|_-=\widetilde{\phi}_j,  &\mbox \qquad \text{on}~\partial B, \\
%		\left.v_j\right|_+-\left.v_j\right|_-=0, &\mbox \qquad \text{on}~\partial B,
%		\end{dcases}
%		\end{equation}
%		and is subject to the Sommerfeld radiation condition, where the eigenvectors $\{\widetilde{\phi}_j\}_{j\in\mathbb{N}}$ form an orthonormal basis of $\mathcal{H}^*(\partial B)$.
%	\end{remark}

	For each normalised eigenfunction of $\mathcal{K}^*_B$, we consider the corresponding function on $\partial D$, 
	\begin{equation*}
	\phi_j(x):=\widetilde{\phi}_j\left(\frac{x-z}{\delta}\right).
	\end{equation*}
	Here $\{\phi_j\}_{j\in\mathbb{N}}$ are the rescaled non-normalised eigenfunctions of $\mathcal{K}_D^*$. Let us introduce
	\begin{equation*}
	\varphi_j:=\frac{\phi_j}{||\phi_j||_{\mathcal{H}^*(\partial D)}}.
	\end{equation*}
	Since $||\widetilde{\phi}_j||_{\mathcal{H}^*(\partial B)}=1$,
	we have (see appendix~\ref{sec:scaling})
	\[
	\varphi_j = 
	\begin{cases}
	\delta^{-1} \phi_j, & d=2,\\ 
	\delta^{-3/2}\phi_j, & d=3.
	\end{cases}
	\]
	Going back to the original unscaled problem:
	\begin{proposition} \label{prop:spec-decomp}
		As $\omega\delta c^{-1}\ll 1$, the spectral decomposition of the field is as follows
		\begin{align}\label{eq:spec-decomp}
		u(x) =\left\{\begin{aligned} & \sum_{j=0}^{J}\frac{1}{\tau_j(\omega)}\left\langle F,\varphi_j\right\rangle_{\mathcal{H}^*(\partial D)}\mathcal{S}^{k_m}_D[\varphi_j](x) + u^\text{in}(x),  &\mbox{} x \in \mathbb{R}^d  \setminus \overline{D},\\ 
		& \sum_{j=0}^{J}\frac{1}{\tau_j(\omega)}\left\langle F,\varphi_j\right\rangle_{\mathcal{H}^*(\partial D)}\mathcal{S}^{k_c}_D[\varphi_j](x), &\mbox{} x \in D.\ \end{aligned}\right.
		\end{align}
	\end{proposition} 
	
	\begin{proof}
	The scaling lemma~\ref{lem:scaling_1} gives $\mathcal{S}^{k_m\delta}_B[\widetilde{\phi}_j](X)=\delta^{-1}\mathcal{S}^{k_m}_D[\phi_j](x)$ for $d=2,3$. From lemma~\ref{lem:scaling_2}, we have $\langle \widetilde{F},\widetilde{\phi}_j \rangle_{\mathcal{H}^*(\partial B)}=\delta^{-3}\left\langle F,\phi_j \right\rangle_{\mathcal{H}^*(\partial D)}$ for $d=3$ and $\langle \widetilde{F},\widetilde{\phi}_j \rangle_{\mathcal{H}^*(\partial B)}=\delta^{-2}\left\langle F,\phi_j \right\rangle_{\mathcal{H}^*(\partial D)}$ for $d=2$.	
	\end{proof}

\section{Plasmonic resonances}\label{sec:res}

\subsection{Size dependant resonant frequencies}
In this section we calculate size and frequency dependent plasmonic resonances. Let $j\in \{0,..,J\}$. Recall that
	\begin{equation*}
	\tau_j(\omega)=
	\begin{dcases}
	\tau_j+\left(\omega \delta c^{-1}\right)^2 \log{\left(\omega \delta c^{-1}\right)}\tau_{j,1}+\mathcal{O}\left(\left(\omega \delta c^{-1}\right)^2\right), & d=2, \\
	\tau_j+\left(\omega \delta c^{-1}\right)^2\tau_{j,2}+\mathcal{O}\left(\left(\omega \delta c^{-1}\right)^3\right), & d=3.
	\end{dcases}
	\end{equation*}

	\begin{definition}
		We say that $\omega$ is a static plasmonic resonance if $\left|\tau_j\right|=0$.
	\end{definition}
	
	\begin{definition}
		We say that $\omega$ is first-order corrected plasmonic resonance if $$\left| \tau_j+(\omega\delta c^{-1})^2\log{(\omega\delta c^{-1})}\tau_{j,1}\right| =0 $$ or $$\left| \tau_j+(\omega\delta c^{-1})^2\tau_{j,2}\right| =0,$$ with $d=2$ or $d=3$, respectively.
	\end{definition}

\begin{remark}
For $j=0$, we have $\tau_0=1/\varepsilon_m$, which is of size one by assumption. We exclude $j=0$ from the set of resonances.
\end{remark}	

For $j\geq 1$ we have $\mathcal{P}_{\mathcal{H}_0^*}[\widetilde{\phi}_j]=\widetilde{\phi}_j$. Let us define
	\begin{equation*}
		\alpha_j:=
	\begin{dcases}
\left\langle \left(\frac{1}{2}I-\mathcal{K}^*_B\right)\widetilde{\mathcal{S}}_B^{-1}\mathcal{S}_{B,1}^{(1)}[\widetilde{\phi}_j], \widetilde{\phi}_j \right\rangle_{\mathcal{H}^*(\partial B)} , & d=2, \\
	\left\langle \left(\frac{1}{2}I-\mathcal{K}^*_B\right)\mathcal{S}_B^{-1}\mathcal{S}_{B,2}[\widetilde{\phi}_j], \widetilde{\phi}_j \right\rangle_{\mathcal{H}^*(\partial B)} , & d=3.
	\end{dcases}
	\end{equation*}
Then, we can calculate
	\begin{equation*}
	\tau_j(\omega)=
	\begin{dcases}
	\frac{\varepsilon_m-\varepsilon_c}{\varepsilon_m\varepsilon_c}\left(\lambda(\omega)-\lambda_j+\left(\omega\delta c^{-1}\right)^2\log{\left(\omega\delta c^{-1}\right)}\alpha_j\right)+\mathcal{O}\left(\left(\omega \delta c^{-1}\right)^2\right), & d=2, \\
	\frac{\varepsilon_m-\varepsilon_c}{\varepsilon_m\varepsilon_c}\left(\lambda(\omega)-\lambda_j+\left(\omega\delta c^{-1}\right)^2\alpha_j\right)+\mathcal{O}\left(\left(\omega \delta c^{-1}\right)^3\right), & d=3.
	\end{dcases}
	\end{equation*}
	
		\begin{lemma}
We have $\alpha_j \in \R$ and
		\begin{equation*}
		\alpha_j:=
	\begin{dcases}
\left(\lambda_j-\frac{1}{2} \right) \left\langle \mathcal{S}_{B,1}^{(1)} [\widetilde{\phi}_j], \widetilde{\phi}_j\right\rangle_{-1/2,1/2} , & d=2,\\
	\left(\lambda_j-\frac{1}{2} \right) \left\langle \mathcal{S}_{B,2} [\widetilde{\phi}_j], \widetilde{\phi}_j\right\rangle_{-1/2,1/2}, & d=3.
	\end{dcases}
	\end{equation*}
	
	\end{lemma}
	
	In what follows we use the lower-case character $\omega$ for real frequencies and the upper-case character $\Omega$ for complex frequencies.

\begin{proposition}\label{prop:reso_3D} 
Using the Drude model \eqref{eq:drude}, the three-dimensional first-order corrected plasmonic resonances $\Omega_j^\pm(\delta):=\pm\Omega'_j+i\Omega''_j$ all lie in the lower part of the complex plane and their modulus is bounded. In the case where we take the medium to be vacuum, i.e., $\varepsilon_m=\varepsilon_0$ we obtain explicitly for $|\lambda_j+1/2| > 10^{-2}$ (this occurs, for example, when $B$ is a ball \cite{photonic}):
		
\begin{equation*}
\Omega_j' =  \sqrt{\frac{\omega_p^2(\lambda_j+1/2)}{1+(\omega_p\delta c^{-1})^2\alpha_j} - \frac{ \mathrm{T}^{-2}}{4\left[1+\left(\omega_p \delta c^{-1}\right)^2 \alpha_j\right]^2}}\quad \text{and }\quad  \Omega_j'' = -\frac{\mathrm{T}^{-1}}{2\left[1+\left(\omega_p \delta c^{-1}\right)^2 \alpha_j\right]}.
\end{equation*}
Moreover, they are bounded
		\begin{equation*}
	|\Omega_j| \leq 2\max \left\{\frac{\mathrm{T}^{-1}}{\left|1+\left(\omega_p\delta c^{-1}\right)^2\alpha_j\right|},\frac{\omega_p\sqrt{\lambda_j+1/2}}{\sqrt{\left|1+\left(\omega_p\delta c^{-1}\right)^2\alpha_j\right|}}\right\}.
		\end{equation*}
\end{proposition}

\begin{proof}
We have that $\tau_j\left(\Omega_j\right) = 0$ if and only if
		\[
		\frac{\Omega_j^2+i\Omega_j \mathrm{T}^{-1}}{\omega_p^2} -\frac{1}{2} - \lambda_j + \frac{\Omega_j^2\delta^2}{c^2} \alpha_j=0,
		\]
		that is
		\[
		\begin{cases}
		\displaystyle \frac{1}{\omega_p^2}\left(\Omega_j'^2-\Omega_j''^2\right) - \frac{1}{2} - \lambda_j + \left(\Omega_j'^2-\Omega_j''^2\right) \frac{\delta^2}{c^2} \alpha_j - \frac{1}{\omega_p^2\mathrm{T}}\Omega_j''^2=0, \\
		\displaystyle \frac{2}{\omega_p^2}\Omega_j' \Omega_j'' + 2\Omega_j' \Omega_j''\frac{\delta^2 }{c^2} \alpha_j + \frac{1}{\omega_p^2 \mathrm{T}} \Omega_j' =0. 
		\end{cases}
		\]
Because $\delta\omega_p c^{-1} \ll 1$, we get the desired result. Lagrange improved upper-bound for roots of polynomials concludes the proof \cite{lagrange}.

		\end{proof}

\begin{definition} \label{def:range_3d}
In three dimensions, we define the resonance radius as
\begin{equation*}
\mathcal{R}(\delta):=\max_{j\in \{1,..,J\}}\max \left\{\frac{2\mathrm{T}^{-1}}{\left|1+\left(\omega_p\delta c^{-1}\right)^2\alpha_j\right|},\frac{2\omega_p\sqrt{\lambda_j+1/2}}{\left|1+\left(\omega_p\delta c^{-1}\right)^2\alpha_j\right|^{1/2}}\right\}.
\end{equation*}
\end{definition}

\begin{remark}
This resonance radius gives our method a range of validity. We compute resonant frequencies in a perturbative quasistatic regime. So  by checking that \begin{align*}
\mathcal{R}(\delta) \delta c^{-1} < \frac{1}{2},
\end{align*} we ensure that the largest plasmonic frequency lies in a region that is still considered as \emph{low-frequency} for a particle of size $\delta$. 
If we pick the size to be too large, namely such that $\mathcal{R}(\delta) \delta c^{-1}$ is bigger than one, it means that the method is not self-consistent, as the largest resonant frequency might not satisfy the $\omega \delta c^{-1} <1/2$.
\end{remark}

\begin{proposition}\label{prop:reso_2D}
In vacuum, and using the Drude model \eqref{eq:drude}, the two-dimensional first-order corrected plasmonic resonances are the roots $\left(\Omega_j\right)_{1\leq j\leq J}\in \C$ of the following equation
\begin{equation}\label{eq:2d_reso_shifted}
\frac{\Omega_j^2+i\Omega_j \mathrm{T}^{-1}}{\omega_p^2} -\frac{1}{2} - \lambda_j +\left(\Omega_j\delta c^{-1}\right)^2 \log{\left(\Omega_j\delta c^{-1}\right)} \alpha_j=0.
\end{equation}
\end{proposition}

%The transcendental equation in Proposition~\ref{prop:reso_2D} is solved numerically using M\"uller's method \cite[Chapter 1]{photonic} and the resonances are plotted in Figure~\ref{fig:reso_2D}. They are also shown to lie in the lower part of the complex plane and their modulus is bounded.

\begin{remark}
We can compute an approximation of the roots of \eqref{eq:2d_reso_shifted} by computing in the first place the static resonances $\left(\Omega_{s,j}\right)_{1\leq j\leq J}$. Solving $\tau_j=0$ yields
\begin{equation*}
\Omega^\pm_{s,j}=\pm \sqrt{\omega_p^2\left(\lambda_j+\frac{1}{2}\right) - \frac{1}{4T^2}} - \frac{i}{2T}.
\end{equation*}
Replacing the dynamic frequency in the logarithm by its static approximation, we transform \eqref{eq:2d_reso_shifted} into the quadratic equation
\begin{equation*}
\frac{\Omega_j^2+i\Omega_j \mathrm{T}^{-1}}{\omega_p^2} -\frac{1}{2} - \lambda_j +\left(\Omega_j\delta c^{-1}\right)^2 \log{\left(\Omega_{s,j}\delta c^{-1}\right)} \alpha_j=0.
\end{equation*}
We get
\begin{equation}
\Omega_j^\pm(\delta)=\frac{-i\mathrm{T}^{-1}\pm\sqrt{4\omega_p^2\left(\lambda_j+1/2\right)\left[1+\alpha_j\left(\omega_p\delta c^{-1}\right)^2\log{\left(\Omega^\pm_{s,j}\delta c^{-1}\right)}\right]
}}{2\left[1+\alpha_j\left(\omega_p\delta c^{-1}\right)^2\log{\left(\Omega^\pm_{s,j}\delta c^{-1}\right)}\right]}.
\label{eq:dynamic}
\end{equation}
\end{remark}

\begin{definition} \label{def:range_2d}
In two dimensions, we define the resonance radius as
\begin{equation*}
\mathcal{R}(\delta):=\max_{j\in \{1,..,J\}}\max \left\{\frac{2\mathrm{T}^{-1}}{\left|1+\alpha_j\left(\omega_p\delta c^{-1}\right)^2\log{\left(\Omega^\pm_{s,j}\delta c^{-1}\right)}\right|},\frac{2\omega_p\sqrt{\lambda_j+1/2}}{\left|1+\alpha_j\left(\omega_p\delta c^{-1}\right)^2\log{\left(\Omega^\pm_{s,j}\delta c^{-1}\right)}\right|^{1/2}}\right\}.
\end{equation*}
\end{definition}

\subsection{Plasmonic quasi-normal modes}\label{sec:QNMdef}

Quasi-normal modes are formally defined as solutions of the source-free wave equation \cite{lalanneReview}. Using the representation formula \eqref{eq:scalar_solution}, we can now define, as in the physics literature, plasmonic quasi-normal modes $(e_j^{\pm})_{j\in\N}$ that oscillate at complex frequencies $\Omega_j^{\pm}(\delta)$:
\begin{align}\label{eq:defQNM}
e_j^{\pm}(x)= \left\{\begin{aligned} & \mathcal{S}^{\Omega_j^{\pm}(\delta)c^{-1}}_D[\varphi_j](x), &\mbox{ } x\in \R^d\setminus \overline{D},\\ 
 & \mathcal{S}^{\Omega_j^{\pm}(\delta)\sqrt{\varepsilon_c \mu_m}}_D[\varphi_j](x),&\mbox{ } x\in D.
 \end{aligned}\right.
\end{align}
These $(e_j^{\pm})_{j\in\N}$ solve the source-free Helmholtz equation and satisfy the radiation condition at infinity, but they diverge exponentially fast as $\vert x\vert \rightarrow \infty$.

\begin{remark} In the physics literature (see \cite[equation (1.1)]{lalanneReview} for instance) one can often find representations of the scattered field in the form
\begin{align*}
u(x,\omega) = \sum_j \alpha_j(u^\text{in}, \omega) e_j^\pm(x),
\end{align*} where $\alpha_j$ are \emph{excitation coefficients} depending on the source and independent of the space variable $x$. These representations are problematic for several reasons. The first one is that any representation of this type is not solution to the Helmholtz equation for $\omega \in \C$ as soon as there are two or more modes oscillating at different frequencies.
The second problem is that in these representations, the scattered wave $u-u^\text{in}$ is not in $L^2(\R^d)$ and only compact subspaces of $\R^d$ can be considered. Then, a renormalisation process is necessary for the eigenmodes since they diverge exponentially. 
Even though the study of these modes \emph{individually} can give physical insight to a system (like for example by studying the mode volume quantity \cite{cognee2019mapping}), they cannot be used in frequency domain representation formulae to solve the scattering problem.
\end{remark}

\section{Time domain approximation of the scattered field}\label{sec:time}

In the following section we show that even though they are irrelevant for frequency domain representation, quasi-normal modes can be used to approximate the field in the time domain.
The idea is to get around costly time domain computations by pre-computing the modes of the system and then expressing the response of the system to any source in terms of the modes. In the physics literature (for example \cite[eq. (1.2)]{lalanneReview}) the field in the time domain is expressed under the form
\begin{align}\label{eq:lalannetimedomain}
u(x,t)= \Re\,  \sum_j \beta_j(t) e_j^\pm(x).
\end{align}
The problem with this type of expansions is that if $\vert x \vert$ is big then $e_j^\pm(x)$ is exponentially large and the computation of $u(x,t)$ is not very stable if the modes are pre-computed. 

We will show in this section that it is possible to express the scattered field in the time domain in a similar expansion, but with non-diverging, pre-computable quantities similar to the quasi-normal modes.

\subsection{The three-dimensional case}
Here we state the main result of the paper, theorem \ref{theo:resonanceexpansion3}, and discuss the result.
\subsubsection{The modal approximation}
Let $\Gamma^{k_m}(\cdot,s)$, i.e., the Green's function for the Helmholtz equation introduced in definition~\ref{de:greenfuction}, be the incident wave $u^\text{in}$ in three dimensions.
Given a wideband signal $\widehat{f}:t \mapsto \widehat{f}(t) \in C_0^{\infty}([0,C_1])$, for $C_1>0$, we want to express the time domain response of the electric field to an oscillating dipole placed at a source point $s$. This means that for a fixed $\delta$ we can pick an excitation signal such that most of the frequency content is in the \emph{low frequencies} but large enough to excite the plasmonic resonances. We can pick  $\eta \ll 1$ and $\rho \geq \mathcal{R}(\delta)$ such that
\begin{align*}
\int_{\R \setminus[-\rho,\rho]} \vert f(\omega) \vert^2 \dd \omega \leq \eta,  \\
\frac{\rho \delta}{c} \leq 1, 
\end{align*}
where $f:\omega\mapsto f(\omega)$ is the Fourier transform of $\widehat{f}$. In practice we take $\rho= \mathcal{R}(\delta)$. 
The incident field has the following form in the time domain:
	\begin{equation*}
	\widehat{u}^\text{in}(x,t)=\int_{\mathbb{R}} \Gamma^{\frac{\omega}{c}}(x,s) f(\omega)e^{-i\omega t} \mathrm{d}\omega=\frac{\widehat{f}(t-|x-s|/c)}{4\pi|x-s|}.
	\end{equation*}
The goal of this section is to establish a resonance expansion for the low-frequency part of the scattered electric field in the time domain.
Introduce, for $\rho>0$,  the truncated inverse Fourier transform of the scattered field $u^{\text{sca}}$ given by 
\begin{equation*}
P_\rho\left[u^\text{sca}\right](x,t):=\int_{-\rho}^{\rho} u^{\text{sca}}(x,\omega) e^{-i\omega t} \mathrm{d}\omega.
\end{equation*}

Recall that $z$ is the centre of the resonator and $\delta$ its radius. Let us define $$t_0^\pm(s,x):=\frac{1}{c}\left(|s-z|+|x-z|\pm 2\delta\right)\pm C_1,$$
the time it takes to the wideband signal to reach first the scatterer and then the observation point $x$. The term $\pm 2 \delta/c$ accounts for the maximal timespan spent inside the particle.

Recall the spectral decomposition in the frequency domain (proposition \ref{prop:spec-decomp}) for $x \in \mathbb{R}^3\setminus \overline{D}$:
\begin{equation*}
u^\text{sca}(x,\omega)=\left(u-u^\text{in}\right)(x,\omega)=\sum_{j=1}^{J}\frac{1}{\tau_j(\omega)}\left\langle F,\varphi_j\right\rangle_{\mathcal{H}^*(\partial D)}\mathcal{S}^{\frac{\omega}{c}}_D[\varphi_j](x).
\end{equation*}

\begin{theorem}\label{theo:resonanceexpansion3}
Let $N\in\N$.
For $J\in\N$ large enough, the scattered field has the following form in the time domain for $x\in\R^3\setminus \overline{D}$: 
	\begin{equation}
	P_{\rho}\left[u^\text{sca}\right](x,t)=
	\begin{dcases}
	\mathcal{O}\left(\delta^4 \rho^{-N}\right), &\mbox{ } t\leq t_0^-, \\
	 2\pi i\sum_{j=1}^J C_{\Omega_j^\pm(\delta)} \langle F, \varphi_j \rangle_{\mathcal{H}^*(\partial D)} e_j^\pm(x) e^{-i\Omega_j^\pm(\delta) t}+\mathcal{O}\left(\frac{\delta^4}{t}\rho^{-N}\right), &\mbox{ }t\geq t_0^+.
	\end{dcases}
	\label{eq:resonanceexpansion3} 
	\end{equation}

The complex numbers $\Omega_j^\pm(\delta)$ are the resonant frequencies given by proposition~\ref{prop:reso_3D}. The fields $e_j$ are the classical  quasi-normal modes  defined in section \ref{sec:QNMdef}.  $C_{\Omega_j^\pm(\delta)}$ is a constant depending only on $j$, the size $\delta$ and the model for $\varepsilon_c(\omega)$:
\begin{align*}
C_{\Omega_j^\pm(\delta)}:=\varepsilon_0 \frac{\left({\Omega^\pm_j(\delta)}^2 + i \Omega^\pm_j(\delta) T^{-1} -\omega_p^2\right) }{\left(1+\left(\omega_p\delta c^{-1}\right)^2 \alpha_j\right)\left(\Omega^\pm_j(\delta)-\Omega^\mp_j(\delta)\right)} .
\end{align*}
\end{theorem}
\begin{remark}
The resonant frequencies $\left\{\Omega_j^\pm(\delta)\right\}_{1\leq j\leq J}$ have negative imaginary parts, so theorem~\ref{theo:resonanceexpansion3} expresses the scattered field as the sum of decaying oscillating fields. The imaginary part of $\Omega_j^\pm(\delta)$ accounts for absorption losses in the particle as well as radiative losses. 
\end{remark}

\begin{remark}[about the remainder $\rho$]
	Since for a particle of finite size $\delta$ our expansion only holds for a range of frequencies $\omega$ such that $\omega\delta c^{-1} <1 $, we cannot compute the full inverse Fourier transform and we have a remainder that depends on the maximum frequency that we can use. Nevertheless  that maximum frequency $\rho$ behaves as $c/\delta$ and we can see that the remainder gets arbitrarily small for small particles.  For a completely point-like particle one would get a zero remainder.
\end{remark}

\begin{remark}\label{rem:fourier} If we had access to the full inverse Fourier transform of the field, of course, since the inverse Fourier transform of a function which is analytic in the upper-half plane is \emph{causal} we would find that in the case $t\leq \left(|s-z|+|x-z|-2\delta\right)/c$, $\widehat{u}^{\text{sca}}(x,t)= 0$. Nevertheless, our method gives the resonant frequencies only in the  \emph{low-frequency} regime. Therefore we only have an approximation for the \emph{low-frequency} part of the scattered field, which does not have a compact support in time. 
Nevertheless, as shown in the numerical section \ref{section:exactvsreference}, the low-frequency part of the scattered field is actually a good approximation for the scattered field. There does not seem to be any resonant frequencies for $\omega> \mathcal{R}(\delta)$. This is highly non-trivial and we do not have a mathematical justification for that. Physically though, it can be explained by looking at the Drude model and noting that when $\omega \rightarrow\infty$, $\varepsilon (\omega)\longrightarrow 1$. The metal does not really interact with light at high frequencies.   \end{remark}

\subsubsection{Alternative formulation with non-diverging causal quasi-normal modes}
Even though $\vert e_j^\pm(x)\vert\longrightarrow \infty $ when $\vert x\vert \rightarrow \infty$, no terms diverge in \eqref{eq:resonanceexpansion3}. Indeed we can rewrite:
\begin{align*}
 e_j^\pm(x) e^{-i\Omega_j^\pm(\delta) t} = &  e_j^\pm(x)e^{-i\Omega_j^\pm(\delta) t_0^+} e^{-i\Omega_j^\pm(\delta) (t-t_0)}\\
 =& e_j^\pm(x)e^{-i\Omega_j^\pm(\delta) \left(|s-z|+|x-z|+ 2\delta\right)c^{-1}+ C_1} e^{-i\Omega_j^\pm(\delta) (t-t_0^+)}\\
 =& C_{u^\text{in}, \delta} e_j^\pm(x) e^{-i\Omega_j^\pm(\delta) |x-z|c^{-1}}e^{-i\Omega_j^\pm(\delta) (t-t_0^+)},
\end{align*}
where $C_{u^\text{in}, \delta}$ depends only on the incoming field and the particle size.
We can define the following \emph{causal plasmonic quasi-normal modes} $(E_j^\pm)_{j\in\N}$ at the complex frequency $\Omega_j^\pm(\delta)$: 
\begin{align}\label{eq:defQNMcausal}
E_j^{\pm}(x)= \left\{\begin{aligned} & \mathcal{S}^{\Omega^\pm_j(\delta) c^{-1}}_D[\varphi_j](x)e^{-i\Omega_j^\pm(\delta) |x-z|c^{-1}}, &\mbox{ }x\in \R^d\setminus \overline{D},\\ 
 & \mathcal{S}^{\Omega^\pm_j(\delta)\sqrt{\varepsilon_c \mu_m}}_D[\varphi_j](x), &\mbox{ } x\in D.
 \end{aligned}\right.
\end{align}
\begin{remark} When referring to $E_j^\pm$, the term \emph{mode} is inaccurate, as $E_j^\pm$ does not solve the Helmholtz equation. But since the $(E_j^\pm)_{j\in\N}$ are built from modes with a complex phase correction, we still call them modes in a loose sense of the term.
\end{remark}
Theorem \ref{theo:resonanceexpansion3} can be re-stated:
\begin{theorem}[alternative causal expansion]\label{theo:resonanceexpansion3-alternate}
\begin{equation}
	P_{\rho}\left[u^\text{sca}\right](x,t)=
	\begin{dcases}
	\mathcal{O}\left(\delta^4 \rho^{-N}\right), &\mbox{ } t\leq t_0^-, \\
	 2\pi i\sum_{j=1}^J \beta_j^{\delta}(u^\text{in})  E_j^\pm(x) e^{-i\Omega_j^\pm(\delta) (t-t_0^+)}+\mathcal{O}\left(\frac{\delta^4}{t}\rho^{-N}\right), &\mbox{ }t\geq t_0^+,
	\end{dcases}
	\label{eq:resonanceexpansion3-alternate} 
	\end{equation}
	where $ \beta_j^{\delta}(u^\text{in})  = C_{\Omega_j^\pm(\delta)} \langle F, \varphi_j \rangle_{\mathcal{H}^*(\partial D)}e^{-i\Omega_j^\pm(\delta) \left(|s-z|+ 2\delta\right)c^{-1}+ C_1}.$
\end{theorem}

\begin{remark} Expansion \eqref{eq:resonanceexpansion3-alternate} has exactly the same form as the representation formula found in the physics literature (like equation \eqref{eq:lalannetimedomain}) but without any exponentially diverging quantities. The $E_j^\pm$ can be computed independently of the source, just like regular quasi-normal modes.
\end{remark}

\subsection{Proof of theorem \ref{theo:resonanceexpansion3}}
Before we can prove theorem \ref{theo:resonanceexpansion3} we need the following lemma:
\begin{lemma}\label{lem:lemma_F}
As $\omega\delta c^{-1} \rightarrow 0$, $F$ defined in \eqref{eq:F_scalar} admits the following asymptotic expansion: 
	\begin{equation} 
	F(x) =
	\frac{1}{\delta}\left[\delta\left(\frac{1}{\varepsilon_c}-\frac{1}{\varepsilon_m}\right)\nu_x\cdot\nabla\Gamma^{k_m}(z-s)+\mathcal{O}\left(\left(\omega\delta c^{-1}\right)^2\right)\right], \quad x \in \partial D.
	\end{equation}
\end{lemma}
\begin{proof}
	See appendix~\ref{subsec:exp_f}.
\end{proof}

\begin{proof}[Proof of theorem \ref{theo:resonanceexpansion3}]
	We start by studying the time domain response of a single mode to a causal excitation at the source point $s$. According to proposition~\ref{prop:spec-decomp} we need to compute the contribution $\Xi_j$ of each mode, that is,  
	\begin{equation*}
	\int_{-\rho}^{\rho}\Xi_j(x,\omega) e^{-i\omega t}\dd \omega :=\int_{-\rho}^\rho\frac{1}{\lambda(\omega)-\lambda_j(\omega \delta)}\left\langle \nabla\Gamma^{\frac{\omega}{c}}(z ,s)\cdot \nu(\cdot) f(\omega),\varphi_j \right\rangle_{\mathcal{H}^*(\partial D)} \mathcal{S}_D^{\frac{\omega}{c}}\left[\varphi_j\right]e^{-i\omega t}\dd \omega,
	\end{equation*}
	where $\lambda_j(\omega \delta) := \lambda_j - \left(\omega\delta c^{-1}\right)^2 \alpha_{j} +\mathcal{O}\left(\left(\omega\delta c^{-1}\right)^3\right)$.
	One can then write:
	\begin{multline*}
	\left\langle \nabla\Gamma^{\frac{\omega}{c}}(z ,s)\cdot \nu(\cdot) f(\omega),\varphi_j \right\rangle_{\mathcal{H}^*(\partial D)} \mathcal{S}_D^{\frac{\omega}{c}}\left[\varphi_j\right]= \\ 
	f(\omega) \left(\frac{1}{|z-s|}-i\frac{\omega}{c}\right)\left(\lambda_j-\frac{1}{2}\right) \int_{\partial D\times \partial D} \frac{v\varphi_j(v)\varphi_j(y)}{16\pi^2|x-y||z-s|}e^{i\frac{\omega}{c}(|x-y|+|z-s|)}\dd \sigma(v) \dd \sigma(y),
	\end{multline*}
	where we used $\left\langle \nu,\varphi_j\right\rangle_{\mathcal{H}^*(\partial D)}=(1/2-\lambda_j)\left\langle x,\varphi_j\right\rangle_{\frac{1}{2},-\frac{1}{2}}$ \cite{ammari2017mathematicalscalar}. Since 
	$\left\langle \nu,\varphi_0\right\rangle_{\mathcal{H}^*(\partial D)}=0$, the zeroth term vanishes in the summation.

	Now we want to apply the residue theorem to get an asymptotic expansion in the time domain. Note that:
	\begin{equation*}
	\int_{-\rho}^{\rho} \Xi_j(x,\omega) e^{-i\omega t} \mathrm{d}\omega= \oint_{\mathcal{C}^{\pm}} \Xi_j(x,\Omega) e^{-i\Omega t}\mathrm{d}\Omega -\int_{\mathcal{C}_\rho^{\pm}} \Xi_j(x,\Omega) e^{-i\Omega t}  \mathrm{d}\Omega,
	\end{equation*}
	where the integration contour $\mathcal{C}_\rho^{\pm}$ is a semicircular arc of radius $\rho$ in the upper (+) or lower (-) half-plane, and $\mathcal{C}^{\pm}$ is the closed contour $\mathcal{C}^{\pm}=\mathcal{C}_\rho^{\pm}\cup[-\rho,\rho]$. 
	The integral on the closed contour is the main contribution to the scattered field by the  mode and can be computed using the residue theorem to get, for $\rho\geq \Re[\Omega_j^\pm(\delta)]$,
	\begin{align*}
	\oint_{\mathcal{C}^{+}}  \Xi_j(x,\Omega) e^{-i\Omega t} \mathrm{d}\Omega&=0,\\
	\oint_{\mathcal{C}^{-}}  \Xi_j(x,\Omega) e^{-i\Omega t} \mathrm{d}\Omega&=2\pi i\text{Res}\left( \Xi_j(x,\Omega)e^{-i\Omega t},\Omega_j^\pm(\delta)\right).
	\end{align*}
	Since $\Omega_j^\pm(\delta)$ is a simple pole of $\omega \mapsto \dfrac{1}{\lambda(\omega)-\lambda_j(\omega \delta)}$ we can write:
	\begin{align*}
	\oint_{\mathcal{C}^{-}} \Xi_j(x,\Omega) e^{-i\Omega t} \mathrm{d}\Omega&=2\pi i\text{Res}\left(\Xi_j(x,\Omega),\Omega_j^\pm(\delta)\right)e^{-i\Omega_j^\pm(\delta)t}.
	\end{align*}
	To compute the integrals on the semi-circle, we introduce:
	\begin{equation*}
	B_j(y,v,\Omega)=\frac{\lambda_j-1/2}{\lambda(\omega)-\lambda_j(\delta\Omega)}\left(\frac{1}{|z-s|}-i\frac{\omega}{c}\right) \int_{\partial D\times \partial D} \frac{v\varphi_j(v)\varphi_j(y)}{16\pi^2|x-y||z-s|} \qquad (y,v) \in (\partial D)^2.
	\end{equation*}
	
	Note that $B_j(\cdot,\cdot,\Omega)$ behaves like a polynomial in $\Omega$ when $\vert \Omega\vert \rightarrow \infty$.
	Given the regularity of the input signal $\widehat{f} \in C_0^{\infty}([0,C_1])$, the Paley-Wiener theorem~\cite[p. 161]{Yosida1995FA} ensures decay properties of its Fourier transform at infinity. For all $N\in\mathbb{N}^*$ there exists a positive constant $C_N$ such that for all $\Omega \in \mathbb{C}$
	\begin{equation*}
	|f(\Omega)|\leq C_N (1+|\Omega|)^{-N}e^{C_1 |\Im{(\Omega)}|}.
	\end{equation*}
	Let $T:=(\vert x-y\vert + \vert s-v\vert)/c$. We now re-write the integrals on the semi-circle
	\begin{align*}
	\int_{\mathcal{C}_\rho^{\pm}} \Xi_j(x,\Omega) e^{-i\Omega t} \mathrm{d}\Omega=\int_{\mathcal{C}_\rho^{\pm}}f(\Omega) \int_{\partial D\times \partial D} B_j(y,v,\Omega) e^{i \Omega \left(T -t\right)}\dd \sigma(v)\dd \sigma(y) \dd \Omega.
	\end{align*}
   We have that $t_0^-+C_1 \leq T \leq t_0^+-C_1 $. Two cases arise. 
	\paragraph{Case 1:}
	For $0<t<t_0^-$ , i.e., when the signal emitted at $s$ has not reached the observation point $x$, we choose the upper-half integration contour $\mathcal{C}^+$. Transforming into polar coordinates, $\Omega=\rho e^{i\theta}$ for $\theta \in [0,\pi]$, we get:
	\begin{align*}
	\left\vert  e^{i \Omega \left(T -t\right)}\right\vert \leq  e^{ -(t_0^--t+C_1)\Im(\Omega)} \qquad \forall (y,v)\in (\partial D)^2,
	\end{align*}
	and
	\begin{align*}
	\left|\int_{\mathcal{C}_\rho^{+}}  \Xi_j(x,\Omega) e^{-i\Omega t} \dd \Omega\right| & \leq \int_0^\pi \rho \left|f\left(\rho e^{i\theta}\right)\right|e^{-\rho (t_0^--t+C_1)\sin{\theta}}\int_{\partial D \times \partial D}\left\vert B_j\left(y,v,\rho e^{i\theta}\right) \right\vert\dd \sigma(v) \dd \sigma(y) \mathrm{d}\theta,\\
	&\leq \rho C_N(1+\rho)^{-N}\delta^4 \max_{\theta\in [0,\pi]}{\left\Vert B_j\left(\cdot, \cdot, \rho e^{i\theta}\right) \right\Vert_{L^{\infty}(\partial D\times \partial D)}} \pi \frac{1-e^{-\rho(t^-_0-t)}}{\rho(t^-_0-t)},
	\end{align*}
	where we used that for $\theta \in [0,\pi/2]$, we have $\sin{\theta} \geq 2\theta/\pi \geq 0$ and $-\cos{\theta}\leq-1+2\theta/\pi$. The usual way to go forward from here is to take the limit $\rho \rightarrow \infty$, and get that the limit of the integral on the semi-circle is zero. However, we work in the quasi-static approximation here, and our modal expansion is not uniformly valid for all frequencies. So we have to work with a fixed maximum frequency $\rho$.. Since $N$ can be taken arbitrarily large and that $B_j$ behaves like a polynomial in $\rho$ \emph{whose degree does not depend on $j$}, we get that, uniformly for $j\in [1,J]$:
	\begin{equation*}
	\left|\int_{\mathcal{C}_\rho^{+}} \Xi_j(x,\Omega) e^{-i\Omega t} \mathrm{d}\Omega\right| = \mathcal{O}\left(\frac{\delta^4}{t_0^--t} \rho^{N}\right).
	\end{equation*}
	Of course if one has to consider the full inverse Fourier transform of the scattered electromagnetic field, by causality, one should expect the limit to be zero. However, one would need high-frequency estimates of the electromagnetic field, as well as a modal decomposition that is uniformly valid for all frequencies. 
	Since our modal expansion is only valid for a limited range of frequencies we get an error bound that is arbitrarily small if the particle is arbitrarily small, but not strictly zero. 
	
	\paragraph{Case 2:}
	For $t>t_0^+$, we choose the lower-half integration contour $\mathcal{C}^-$. Transforming into polar coordinates, $\Omega=\rho e^{i\theta}$ for $\theta \in [\pi,2\pi]$, we get
	\begin{align*}
	\left\vert  e^{i \Omega \left(T -t\right)}\right\vert \leq  e^{ (t-t_0^+-C_1) \Im (\Omega)} \qquad \forall (y,v)\in (\partial D)^2,
	\end{align*}
	and
	%		\begin{align*}
	%		\left|\int_{\mathcal{C}_\rho^{-}} \Xi_n(x,\Omega) e^{-i\Omega t} \mathrm{d}\Omega\right| & \leq \int_\pi^{2\pi}\rho \left|f\left(\rho e^{i\theta}\right)\right|e^{\rho (t-t_0^+-C_1)\sin{\theta}}\int_{\partial D\times \partial D}\left\vert B_n\left(y,v,\rho e^{i \theta}\right) \right\vert\dd \sigma(v) \dd \sigma(y) \mathrm{d}\theta,\\
	%		&\leq \rho C_N(1+\rho)^{-N} \delta^4 \max_{\theta\in [\pi,2\pi]}{\left\Vert B_n\left(\cdot, \cdot, \rho e^{i\theta}\right) \right\Vert_{L^{\infty}(\partial D\times \partial D)}}\pi \frac{1-e^{\rho( C_1-(t-t_0^+))}}{\rho( C_1-(t-t_0^+))},
	%		\end{align*}
	\begin{align*}
	\left|\int_{\mathcal{C}_\rho^{-}} \Xi_j(x,\Omega) e^{-i\Omega t} \mathrm{d}\Omega\right| & \leq \int_\pi^{2\pi}\rho \left|f\left(\rho e^{i\theta}\right)\right|e^{\rho (t-t_0^+-C_1)\sin{\theta}}\int_{\partial D\times \partial D}\left\vert B_j\left(y,v,\rho e^{i \theta}\right) \right\vert\dd \sigma(v) \dd \sigma(y) \mathrm{d}\theta,\\
	&\leq \rho C_N(1+\rho)^{-N} \delta^4 \max_{\theta\in [\pi,2\pi]}{\left\Vert B_j\left(\cdot, \cdot, \rho e^{i\theta}\right) \right\Vert_{L^{\infty}(\partial D\times \partial D)}}\pi \frac{1-e^{-\rho(t-t_0^+)}}{\rho(t-t_0^+)}.
	\end{align*}
	Exactly as in Case $1$, we cannot take the limit $\rho \rightarrow \infty$. Using the fact that $N$ can be taken arbitrarily large and that $B_j$ behaves like a polynomial in $\rho$ \emph{whose degree does not depend on $j$}, we get that, uniformly for $j\in [1,J]$:
	\begin{align*}
	\left|\int_{\mathcal{C}_\rho^{-}} \Xi_j(x,\Omega) e^{-i\Omega t} \mathrm{d}\Omega\right| = \mathcal{O}\left(\frac{\delta^4}{t} \rho^{-N}\right).
	\end{align*}
	The result of theorem~\ref{theo:resonanceexpansion3} is obtained by summing the contribution of all the modes considered. 
\end{proof}

\begin{remark} The fact that we work with a finite number of modes is necessary for the perturbation theory of section \ref{sec:modal_decomp} but also in this section. Indeed, if we consider all the modes there is an accumulation point in the poles of the modal expansion of the field, and therefore we cannot apply the residue theorem. 
\end{remark}

\subsection{The two-dimensional case} \label{subsec:time-domain-2D} 
In two dimensions, the Green's function does not have an explicit phase term, so we need to introduce another asymptotic parameter $\epsilon>0$ to be able to use the large argument asymptotics of the Hankel function. Our new truncated inverse Fourier transform  of the scattered field $u^{\text{sca}}$ given by 
\begin{equation*}
P_{\rho,\epsilon}\left[u^\text{sca}\right](x,t)=\int_{-\rho}^{-\epsilon} u^{\text{sca}}(x,\omega) e^{-i\omega t} \dd\omega+\int_{\epsilon}^{\rho}  u^{\text{sca}}(x,\omega) e^{-i\omega t} \dd\omega.
\end{equation*}
This allows us to define a notion of  \emph{far field}. A point $x$ is far from $D$ if $\epsilon\vert x-z\vert c^{-1}\gg 1$. We can now add two additional hypotheses:
\begin{itemize}
\item the source is far away from the particle (or equivalently, the incoming wave is a plane wave)
\item the observation point is far away from the particle. 
\end{itemize}

The incident field has the following form in the time domain:
	\begin{equation}\label{eq:u_in(t)}
	\widehat{u}^{\text{in}}(x,t)=\widehat{f}\left(t-\frac{d\cdot x}{c}\right).
	\end{equation}

Besides these two assumptions and a difference in the order of the remainder, the result in two dimensions is essentially the same as in three dimensions. 

\begin{theorem}\label{theo:resonanceexpansion2}
	Let $N\in \N$. For $J$ large enough the scattered field has the following form in the time domain for $x$ far away from $D$:
	\begin{equation}
	P_{\rho,\epsilon}\left[u^\text{sca}\right](x,t)=
	\begin{dcases}
	\mathcal{O}\left(\delta \rho^{-N}\right), &\mbox{ } t\leq t_0^-, \\
	2\pi i\sum_{j=1}^J C_{\Omega_j^\pm(\delta)} \langle F, \varphi_j \rangle_{\mathcal{H}^*(\partial D)} e_j^\pm(x) e^{-i\Omega_j^\pm(\delta) t}+\mathcal{O}\left(\frac{\delta}{t}\rho^{-N}\right), &\mbox{ }t\geq t_0^+,
	\end{dcases}
	\label{eq:resonanceexpansion2} 
	\end{equation}
	with $\Omega_j^\pm(\delta)$ being the plasmonic resonant frequencies of the particle given by proposition~\ref{prop:reso_2D}.  $C_{\Omega_j^\pm(\delta)}$ is a constant depending only on $j$, the size $\delta$ and the model for $\varepsilon_c(\omega)$:
\begin{align*}
C_{\Omega_j^\pm(\delta)}:=\varepsilon_0 \frac{\left({\Omega^\pm_j(\delta)}^2 + i \Omega^\pm_j(\delta) T^{-1} -\omega_p^2\right) }{\left(1+\left(\omega_p\delta c^{-1}\right)^2 \log \left(\Omega^\pm_{s,j}\delta c^{-1}\right)\alpha_j\right)\left(\Omega^\pm_j(\delta)-\Omega^\mp_j(\delta)\right)} .
\end{align*}
\end{theorem}

\begin{proof}
The proof is quite similar to the three-dimensional case. It is included in appendix \ref{sec:appendixproof2D} for the sake of completeness.
\end{proof}

\section{Numerical simulations}\label{sec:num}

The goal of this section is to illustrate the validity of our approach and to show that the approximation seems to be working with less restrictive hypotheses than the ones in theorem \ref{theo:resonanceexpansion2}:
\begin{itemize}
\item for more general shapes (non-convex or non-algebraic)
\item closer to the particle (outside of the far field approximation).
\end{itemize}
For these simulations we build upon the codes for the layer potentials developed in~\cite{wangcode}.

\subsection{Domains and physical parameters}

\begin{figure}[H]
		\centering
			\includegraphics[trim={0.5cm 6.5cm 0.5cm 6cm},clip,width=15cm]{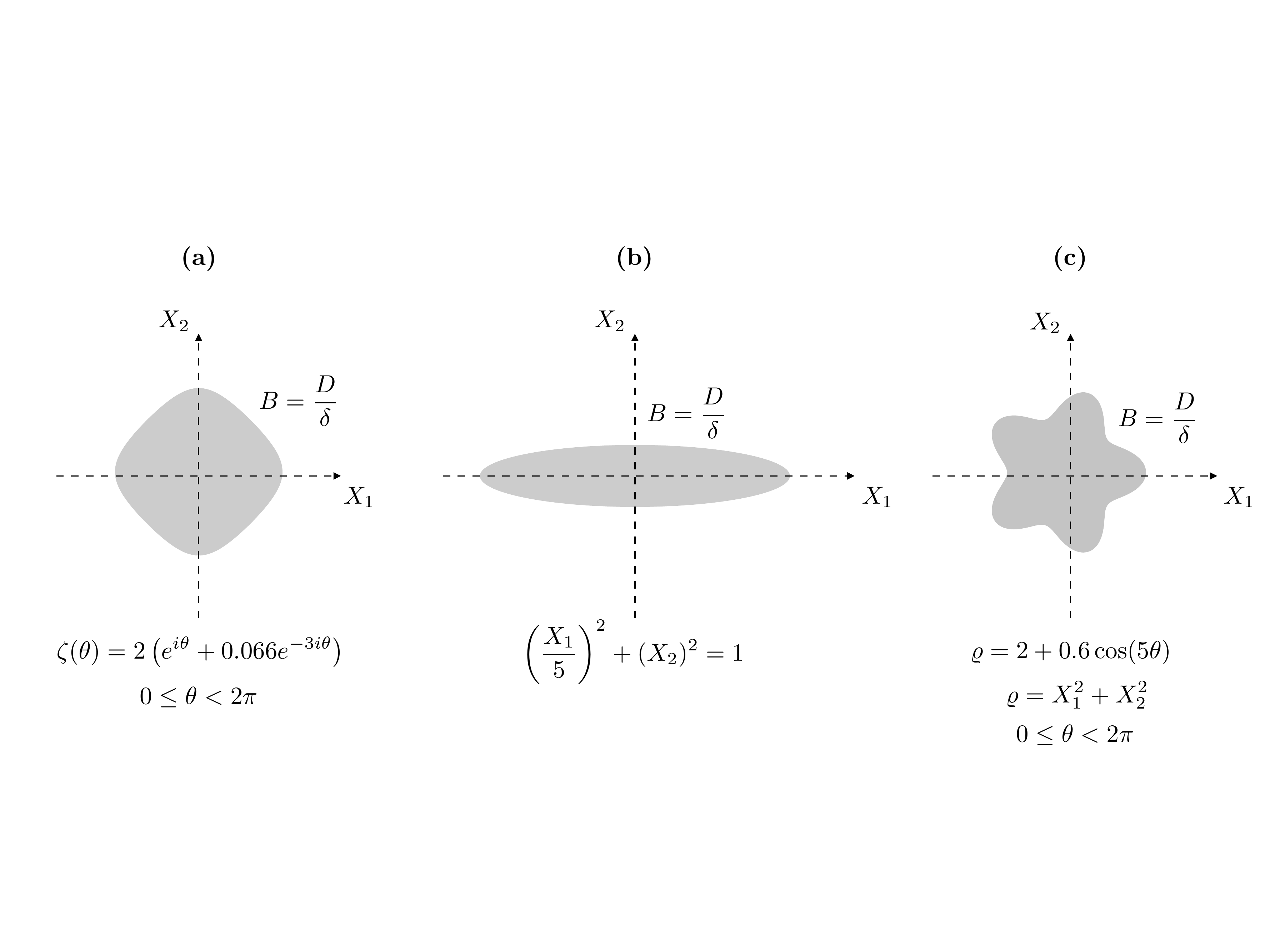} 
		\caption{Sketch of the three reference domains: the rounded diamond (a), the narrow ellipse (b) and the five-petal flower (c).}
		\label{fig:domains}
	\end{figure}

Throughout this section, we consider the three domains sketched on Figure~\ref{fig:domains} to illustrate our results:
\paragraph{Rounded diamond:}

The rounded diamond (a) is defined by the parametric curve $\zeta(\theta)=2\left(e^{i\theta}+0.066e^{-3i\theta}\right)$, for $\theta\in[0,2\pi]$. It is an algebraic domain of class $\mathcal{Q}$ from \cite{putinar19}. This shape satisfies condition~\ref{cond:2d}, as well as the hypotheses of theorem~\ref{theo:resonanceexpansion2}. 

\paragraph{Narrow ellipse:}

The ellipse (b) semi-axes are on the $X_1$- and $X_2$- axes and are of length $a=1$ and $b=5$, respectively. It is algebraic but not asymptotically a circle in the sense of \cite{putinar19}.

\paragraph{Five-petal flower:}

The flower (c) is defined by $\varrho=2+0.6\cos(5\theta)$ in polar coordinates. It has Cartesian equation
\begin{equation*}
0.5\left(X_1^2+X_2^2\right)^3-1.5X_1\left(X_1^2+X_2^2\right)^2+6X_1^3\left(X_1^2+X_2^2\right)-4.8X_1^5\left(X_1^2+X_2^2\right)^3-\left(X_1^2+X_2^2\right)^{5/2}=0
\end{equation*}
 in the rescaled $(X_1,X_2)$ plane. So it is not algebraic (due to the non-integer power of the last term) and not convex. We have no theoretical results on the number of modes that radiate.

%The first four modes $v_j=\mathcal{S}^{k_m\delta}_B[\widetilde{\phi}_j]$ are shown in Figure~\ref{fig:modes_all} for both domains.

All three domains $D=z+\delta B$ are centred at the origin $(z=0)$ for simplicity. We set the size of the nanoparticle to be $\delta=10^{-8}$m. 
%		\begin{figure}[H]
%		\centering
%			\includegraphics[trim={1.4cm 6cm 5cm 9.8cm},clip,width=15cm]{modes_all.pdf} 
%		\caption{We plot the real part of first four modes $v_j=\mathcal{S}^{k_m\delta}_B[\widetilde{\phi}_j]$, $j=1,\ldots,4$ at the wavenumber $k_m=\omega_p/c$ for the ellipse on panel (a) and for the flower on panel (b).}
%		\label{fig:modes_all}
%	\end{figure}	
The numerics are performed on the rescaled domain $B$ and the homogeneous medium is taken to be vacuum ($\varepsilon_m=\varepsilon_0$ and $\mu_m=\mu_0$). The physical parameter values are summarised in Table~\ref{tab:1}.

\begin{table}
	\centering
	\begin{tabular}{|c|c|}
		\hline 
		\textit{Symbol} & \textit{Magnitude} \\
		\hline \hline 
		$\omega_p$  & $2\cdot 10^{15}$ Hz  \\
		\hline
		$T$  & $10^{-14}$s  \\
		\hline
		$\varepsilon_0$  & $8.854187128 \cdot 10^{-12}$ Fm$^{-1}$ \\
		\hline
		$\mu_0$  & $4\pi \cdot 10^{-7}$ Hm$^{-1}$ \\
		\hline
		$\delta$  & $10^{-8}$ m  \\
		\hline 
		d  & $(1/\sqrt{2},1/\sqrt{2})$ \\
		\hline 
		z  & $(0,0)$ \\
		\hline 
	\end{tabular}
	\caption{Physical constants and parameters values. }
	\label{tab:1}
\end{table} 
	
\subsection{Modes contribution decay}	
\begin{figure}[H]
		\centering
			\includegraphics[trim={0cm 4.5cm 0cm 8cm},clip,width=15cm]{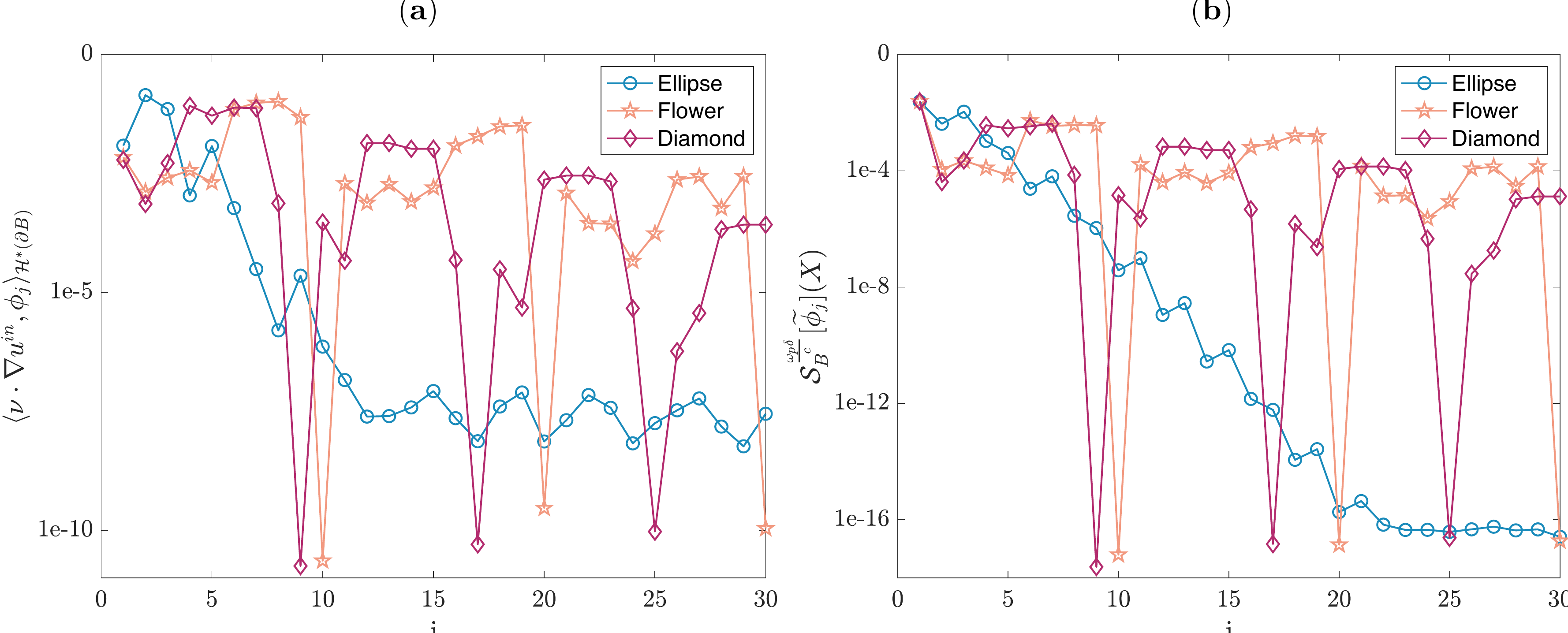} 
		\caption{We illustrate on a logarithmic scale the fast decay of the modal expansion terms by plotting the scalar products $\langle\nu\cdot \nabla u^\text{in},\widetilde{\phi}_j\rangle_{\mathcal{H}^*(\partial B)}$ on panel (a) and the modes $\mathcal{S}^{\omega_p\delta c^{-1}}_B[\widetilde{\phi}_j](X)$ on panel (b), against $j$, for $1\leq j\leq 30$, for the diamond, the ellipse and the flower.}
	\label{fig:truncate_2D}
\end{figure}

It was shown in section \ref{subsec:truncate} that the scalar products $\langle\widetilde{F},\widetilde{\phi}_j\rangle_{\mathcal{H}^*(\partial B)}$ decay very rapidly when $d=3$. In a two-dimensional setting, the theoretical framework is not as clear, but we check numerically that the contribution the modes decrease quite fast with $j$.
Recall that the weight of the $j^\text{th}$ mode is given by the scalar product $\langle\widetilde{F},\widetilde{\phi}_j\rangle_{\mathcal{H}^*(\partial B)}$, which, in a low-frequency regime, can be approximated as $\langle\nu\cdot \nabla u^\text{in},\widetilde{\phi}_j\rangle_{\mathcal{H}^*(\partial B)}$ (see lemma~\ref{lem:lemma_F_2D}). On panel (a) of Figure~\ref{fig:truncate_2D} we show on all examples that $\langle\nu\cdot \nabla u^\text{in},\widetilde{\phi}_j\rangle_{\mathcal{H}^*(\partial B)}$ decays as $j$ grows. We average over all possible directions $d$ of the incident field. Panel (b) of the same picture shows that the modes themselves, $\mathcal{S}^{\omega_p\delta c^{-1}}_B[\widetilde{\phi}_j](X)$, decrease as $j$ increases. We average here over all observation positions, $X$ belongs to a circle of radius $100$ centred at $z=0$. 

\subsection{Plasmonic resonances}
\begin{figure}[H]
		\centering
			\includegraphics[trim={0cm 7cm 0cm 6.2cm},clip,width=15cm]{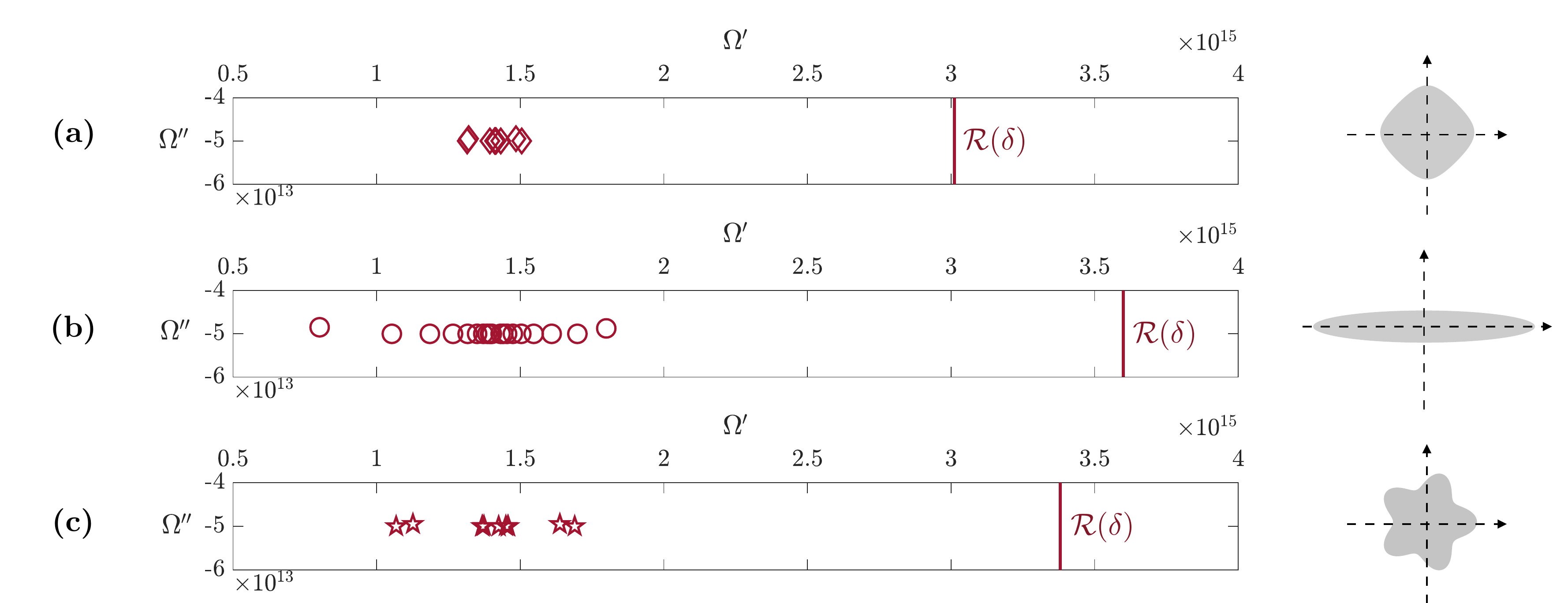} 
		\caption{We plot, for the diamond (a), the ellipse (b) and the flower (c), the two-dimensional first-order corrected resonances with positive real parts from \eqref{eq:dynamic}: $\Omega^+_j (\delta)= \Omega_j' + i \Omega_j''$, for $j=1,..,20$. These resonances lie in the lower part of the complex plane and their real part is between $\omega_p/4$ and $\omega_p$ (and smaller than $\mathcal{R}(\delta)$). Their negative counterparts are symmetric with respect to the imaginary axis.}
	\label{fig:reso_2D}
\end{figure}

We plot the first-order corrected plasmonic resonances with positive real parts on Figure~\ref{fig:reso_2D}. The resonance radius $\mathcal{R}(\delta)$ from definition~\ref{def:range_2d} is drawn as a red vertical line on the three subplots and is shown to encompass all the low-frequency resonances.
We can then verify a posteriori that our choice of size $\delta$ is consistent by checking that $\mathcal{R}(\delta)$ is still in the \emph{low-frequency} region, see table \ref{table:R(delta)}.

\begin{table}
\begin{center}
\begin{tabular}{|c | c | c| }
\hline $B$  & $\mathcal{R}(\delta)$ & $\mathcal{R}(\delta) \delta c^{-1}$ \\ 
\hline \hline Diamond &  $3.0117e+15$ & $0.1005$ \\ \hline
Ellipse &  $3.6228e+15$ & $0.1208$ \\ \hline 
Flower  & $3.3806e+15$ & $0.1128$ \\ \hline 
\end{tabular}
\end{center}
\caption{Validity check}\label{table:R(delta)}
\end{table}

\subsection{Validation of theorem~\ref{theo:resonanceexpansion2}}
In this section, we validate the two-dimensional approximation of the scattered wave in the time domain given in theorem~\ref{theo:resonanceexpansion2} by plotting the asymptotic result against full numerical simulations.

\begin{figure}
		\centering
			\includegraphics[trim={0cm 3.9cm 0cm 5cm},clip,width=15cm]{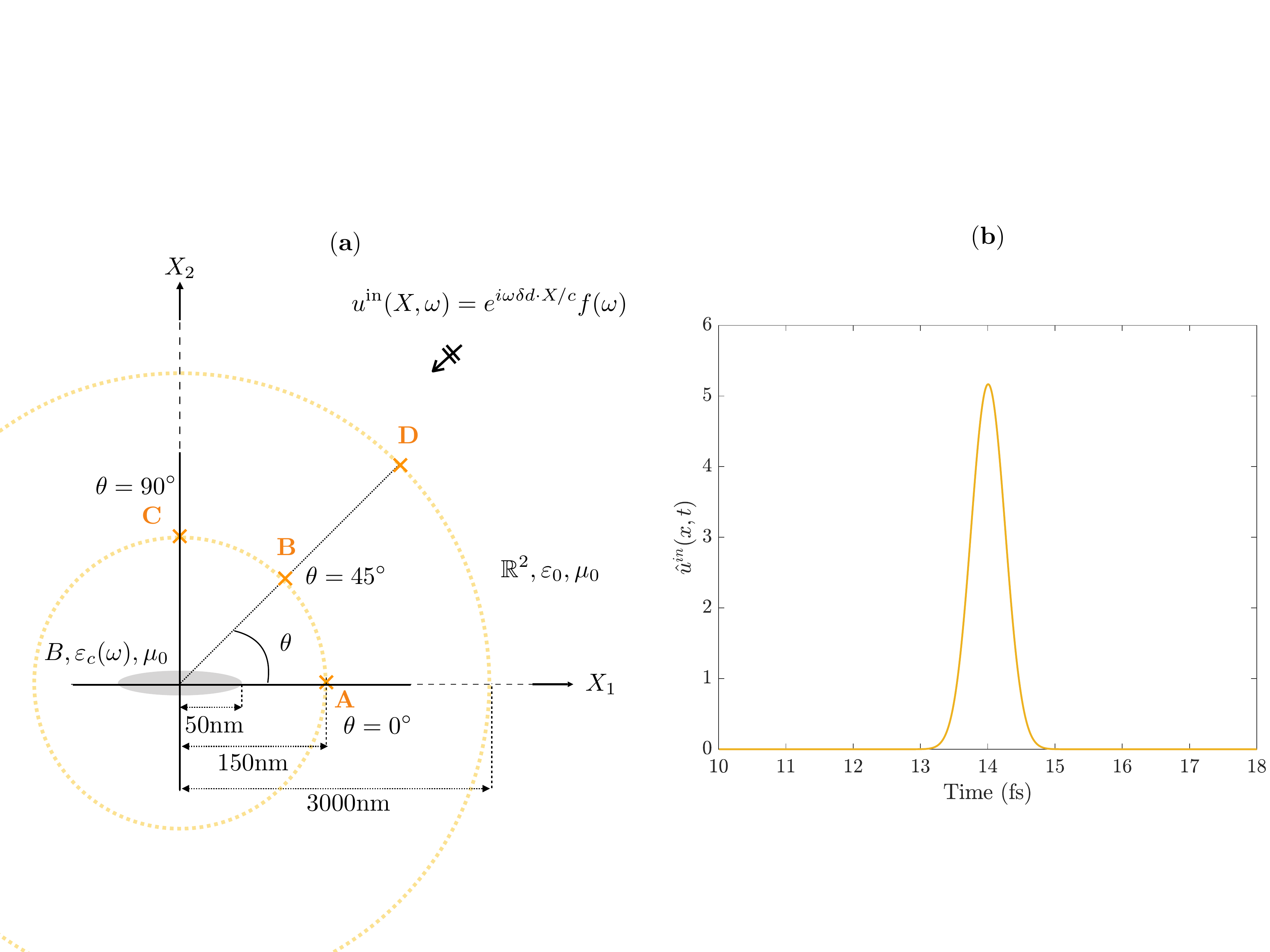} 
		\caption{Not-to-scale sketch of the simulation setting for the ellipse on panel (a). The observation points $A$, $B$ and $C$ are placed on a circle of radius $150$ nm ($|X|=15$) centred at the origin, while observation point $D$ is placed in the far-field on a circle of radius $3000$ nm ($|X|=300$) at angle $\theta_D=45^\circ$. On panel (b) we plot the time domain incident wave $u^\text{in}(x,t)$ from \eqref{eq:u_in(t)} at $x=3000$ nm.}
		\label{fig:setting}
\end{figure}

We sketch the simulation setting with the ellipse in Figure~\ref{fig:setting}(a). We define three observation points $A$, $B$ and $C$ on a circle of radius $150$ nm ($|X|=15$) and one observation point $D$ on a circle of radius $3000$ nm ($|X|=300$). They are characterised by their angle with respect to the x-axis: $\theta_A=0^\circ$, $\theta_B=\theta_D=45^\circ$, $\theta_C=90^\circ$. The nanoparticle is illuminated by a plane wave of the form $u^\text{in}(X)=e^{i k_m d\cdot \delta X} f(\omega)$ where $f$ is the Fourier transform of a bump function compactly supported in the interval $[0,C_1]$, with $C_1= 8$ fs. We plot the time domain incoming wave in Figure~\ref{fig:setting}(b). To ease the notations we drop the tilde subscript in the following and write $u(X)$ instead of $\widetilde{u}(X)$. 

\subsubsection{Reference solution}
We call \emph{reference solution} the low-frequency part of the scattered field in the time domain.
We first uniformly discretise our frequency domain $I_\omega$ in $L=10^4$ points, with  
\begin{align*}
I(\omega):=& [-\rho \delta c^{-1}, -\epsilon \delta c^{-1}] \cup [\epsilon \delta  c^{-1}, \rho \delta c^{-1}] \\ =& [-\omega_p \delta c^{-1}, -\omega_p \delta c^{-1}/4]\cup [\omega_p \delta c^{-1}/4,\omega_p \delta c^{-1}],
\end{align*}
by setting $\omega_{l}$ such that:
$$
-\rho \delta c^{-1}=\omega'_{-L} < \omega'_{-L+1} < \ldots < \omega'_{-1}= -\epsilon \delta c^{-1}, \hspace{3mm} \epsilon \delta  c^{-1} =\omega'_{1} < \ldots < \omega'_{L-1} <  \omega'_{L}=\rho \delta c^{-1},
$$
with $\omega'_{l+1}-\omega'_{l}=(\rho-\epsilon)\delta c^{-1}/L$ for every $l \in [-L-1,-1]\cup [1,L-1]$.
We compute the scattered field in the frequency domain using the representation formula \eqref{eq:scalar_solution}. The single layer potential is approximated using $N=2^8$ equally-spaced discretisation points along the boundary $\partial B$. We define the dimensionless frequency $\omega'=\omega \delta c^{-1}$. The reference solution is computed by taking the truncated inverse Fourier transform 
\begin{equation}\label{eq:reference_sol}
P_{\rho,\epsilon}\left[\widehat{u}^\text{sca}\right](X, t) \approx  \frac{(\rho - \epsilon)}{L} \sum_{l=1}^L \left(e^{-i\omega'_{-l}c \delta^{-1}t}u^\text{sca}\left(X,\omega'_{-l}c \delta^{-1}\right) + e^{-i\omega'_{l}c \delta^{-1}t}u^\text{sca}\left(X,\omega'_{l}c \delta^{-1}\right)\right).
\end{equation} 
%c \delta^{-1} P_{\rho \delta c^{-1},\epsilon \delta c^{-1}}\left[\widehat{u}^\text{sca}\right](X,c\delta^{-1} t) 

\subsubsection{Asymptotic solution}
The expansion is obtained by summing the first $J=30$ modes. Using theorem \ref{theo:resonanceexpansion2}, the modal approximation of order $J$ becomes:

\begin{equation}\label{eq:asymp_sol}
\begin{aligned}
U_J(X,t)=& 2 \pi i \sum_{j=1}^J\varepsilon_0 \left( \frac{\left({\Omega^+_j(\delta)}^2 + i \Omega^+_j(\delta) T^{-1} -\omega_p^2\right) \langle \widetilde{F}, \widetilde{\phi}_j \rangle_{\mathcal{H}^*(\partial B)}}{\left(1+\left(\omega_p\delta c^{-1}\right)^2 \log \left(\Omega^+_{s,j}\delta c^{-1}\right)\alpha_j\right)\left(\Omega^+_j(\delta)-\Omega^-_j(\delta)\right)} \delta \mathcal{S}^{\Omega^+_j(\delta)\delta c^{-1}}_B[\widetilde{\phi}_j](X) e^{-i\Omega^+_j(\delta)t} \right. \\ 
&\left. + \frac{\left({\Omega^-_j(\delta)}^2 + i \Omega^-_j(\delta) T^{-1} -\omega_p^2\right) \langle \widetilde{F}, \widetilde{\phi}_j \rangle_{\mathcal{H}^*(\partial B)}}{\left(1+\left(\omega_p\delta c^{-1}\right)^2 \log \left(\Omega^-_{s,j}\delta c^{-1}\right)\alpha_j\right)\left(\Omega^-_j(\delta)-\Omega^+_j(\delta)\right)} \delta \mathcal{S}^{\Omega^-_j(\delta)\delta c^{-1}}_B[\widetilde{\phi}_j](X) e^{-i\Omega^-_j(\delta)t}\right).
\end{aligned} 
\end{equation}

The simulation results are shown in figures~\ref{fig:diamond_final}, \ref{fig:ellipse_final}, \ref{fig:add_modes} and \ref{fig:flower_final}. To corroborate our pole expansion, we plot the real part of the reference solution \eqref{eq:reference_sol} against the real part of the asymptotic one \eqref{eq:asymp_sol} for the different domains and from different observation points.

\subsubsection{Comparison in the far-field for the diamond}
\begin{figure}[H]
		\centering
			\includegraphics[trim={0cm 6.2cm 0cm 6.7cm},clip,width=15cm]{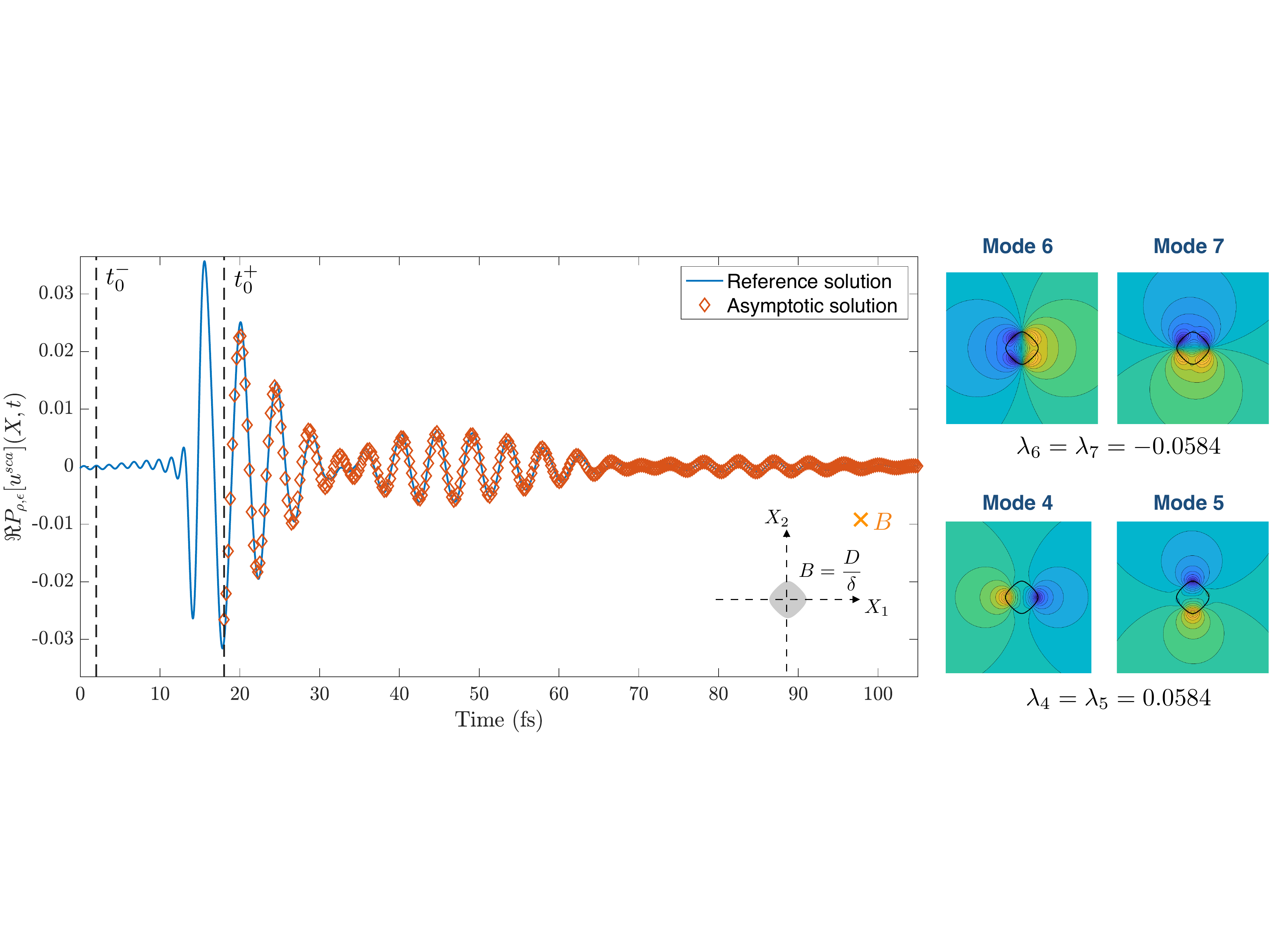} 
		\caption{The real part of the reference solution (blue line) from \eqref{eq:reference_sol} against the real part of the asymptotics (orange symbols) from \eqref{eq:asymp_sol} for the diamond, from observation point $X=D$. The four modes with the largest amplitude are shown on the right (order left to right, up to down).}
		\label{fig:diamond_final}
	\end{figure}

We begin with the diamond, since it is the shape that satisfies the hypotheses of theorem~\ref{theo:resonanceexpansion2}. Figure~\ref{fig:diamond_final} shows the field scattered by the diamond, measured in the far-field at position $X=D$. The reference solution is nicely approximated by the sum of four modes (4, 5, 6 and 7).

\subsubsection{Extension to a nearer-field for the ellipse and flower}
\begin{figure}[H]
		\centering
			\includegraphics[trim={0cm 4.5cm 0cm 7.7cm},clip,width=15cm]{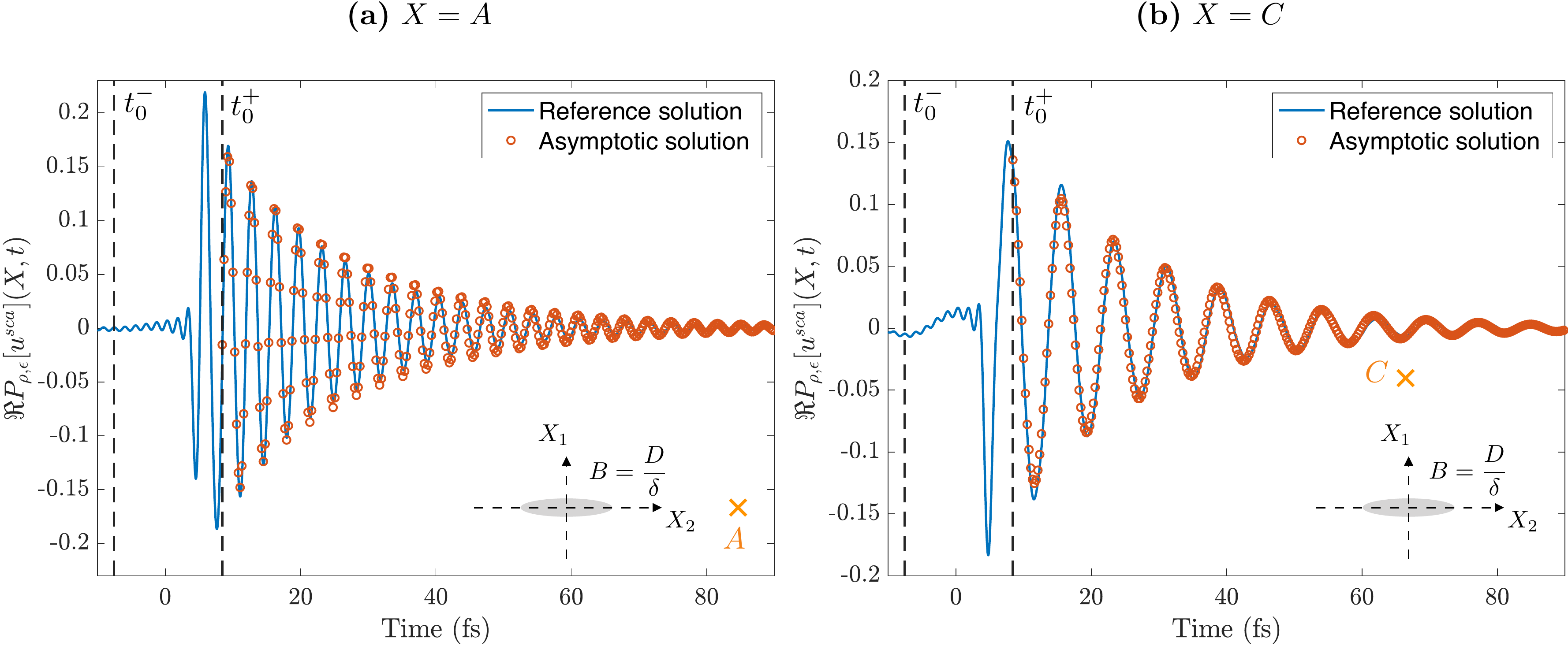} 
		\caption{The real part of the reference solution (blue line) from \eqref{eq:reference_sol} against the real part of the asymptotics (orange symbols) from \eqref{eq:asymp_sol} for the ellipse, from observation point $X=A$ on panel (a) and $X=C$ on panel (b).}
		\label{fig:ellipse_final}
	\end{figure}

Figure~\ref{fig:ellipse_final} shows the field scattered by the ellipse, measured at position $X=A$ on panel (a) and $X=C$ on panel (b). In both cases the time domain scattered wave (blue line) is well approximated by the sum of decaying modes (orange symbols). Although we compute the first $30$ terms of the modal expansion, the actual number of modes which contribute significantly to approximate the reference solution is much smaller. Indeed, only $1$ mode is necessary to reconstruct more than $99\%$ of the signal in Figure~\ref{fig:ellipse_final}. 
	
\begin{figure}[H]
		\centering
			\includegraphics[trim={0cm 0cm 0cm 0cm},clip,width=15cm]{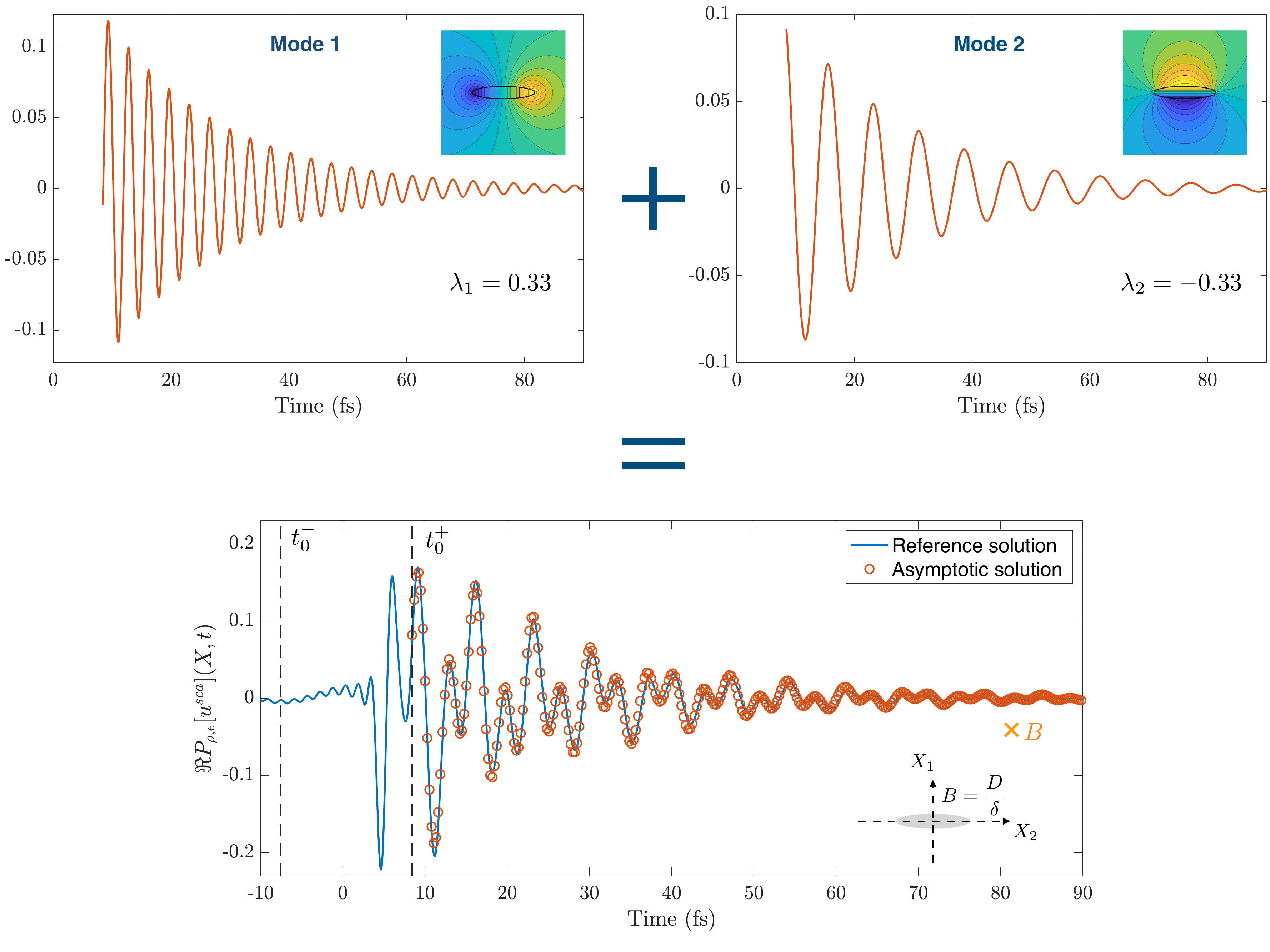} 
		\caption{The real part of the field scattered by the ellipse and observed at point $X=B$ is the superposition of two dipoles modes. The modes (upper panels) oscillate at different frequencies. On the lower panel, the reference solution from \eqref{eq:reference_sol} captures well the expansion from \eqref{eq:asymp_sol}.}
		\label{fig:add_modes}
\end{figure}

When the observation point is at $X=B$, we illustrate in Figure~\ref{fig:add_modes} that two modes are needed to match the reference solution for the ellipse. Mode $1$, corresponding to a dipole which radiates most of the energy along the x-axis, is associated to the eigenvalue $\lambda_1=0.33$. Mode $2$ corresponds to the dipole which radiates most of the energy along the y-axis and is associated to the eigenvalue $\lambda_1=-0.33$. Mode $1$ oscillates slightly faster than mode $2$, resulting in the double oscillation visible on the lower plot. These numerical simulations are in line with \cite{ammari2018finedetails}. Even relatively close to the particle (the observation distance is about a tenth of the wavelength), only two modes radiate in the far-field.

\begin{figure}[H]
		\centering
			\includegraphics[trim={0cm 6cm 0cm 0cm},clip,width=15cm]{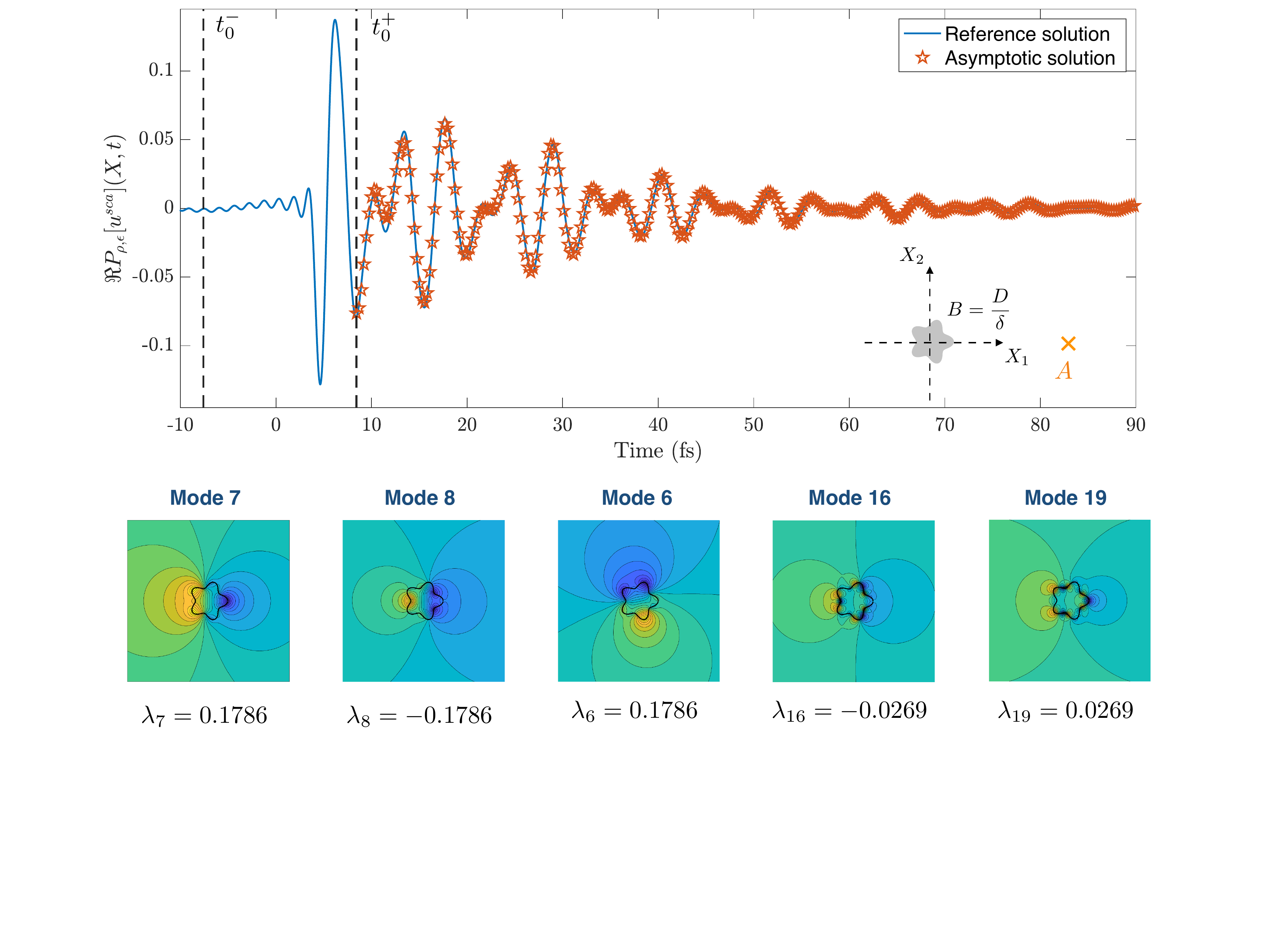} 
		\caption{We plot the real part of the reference solution \eqref{eq:reference_sol} as a blue line against the real part of the asymptotic one \eqref{eq:asymp_sol} as orange symbols for the flower. The observation point is at $X=A$ (shown on the not-to-scale inset). The five modes with the largest amplitude are shown on the bottom (order left to right).}
		\label{fig:flower_final}
	\end{figure}

Figure~\ref{fig:flower_final} shows that even for the non-algebraic flower shape, the scattered wave (blue line) is well approximated by the sum of a small number of decaying modes (orange symbols). As anticipated by Figure~\ref{fig:truncate_2D}, the modes decay being faster for the ellipse than it is for the flower, a larger number of modes is needed for the latter. In Figure~\ref{fig:flower_final}, eight modes were needed to reconstruct more than $99\%$ of the reference solution (and five modes sufficed for $95\%$).

\subsubsection{About the high frequencies}\label{section:exactvsreference}

\begin{figure}[H]
		\centering
			\includegraphics[trim={0cm 11cm 0cm 3cm},clip,width=15cm]{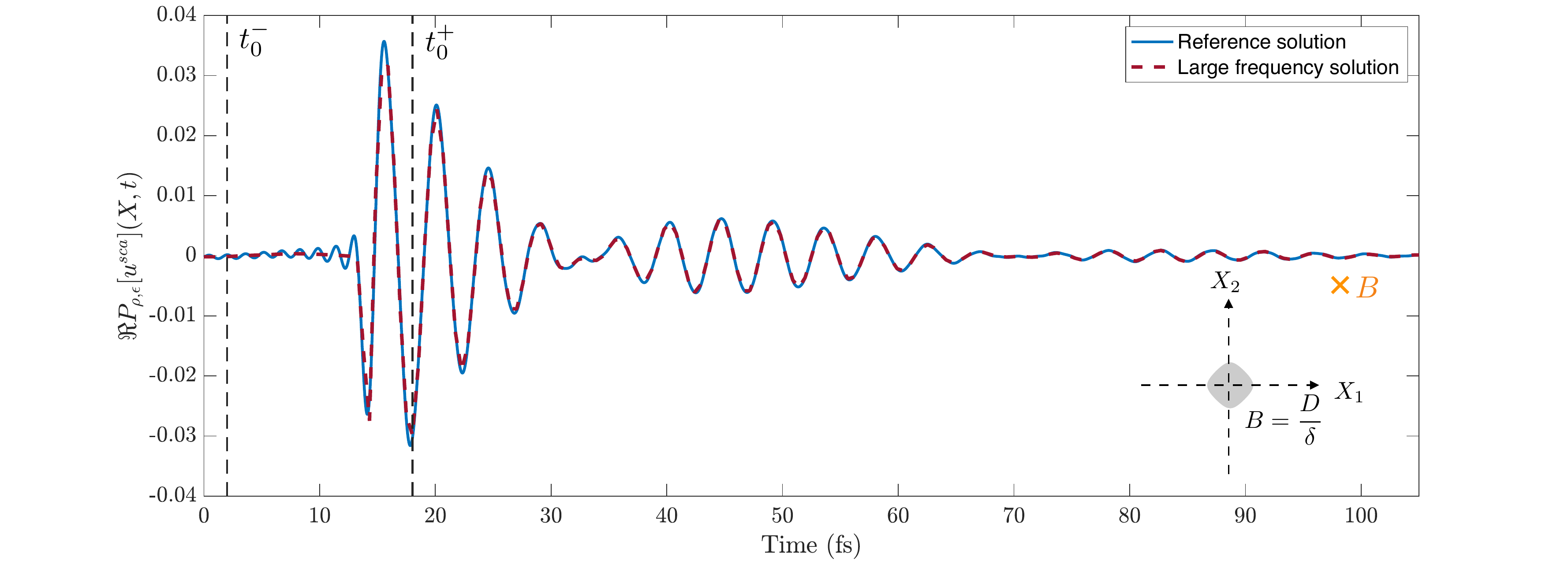} 
		\caption{Reference (low-frequency) solution (computed with $\rho=\mathcal{R}(\delta)$) against large-frequency solution (computed with $\rho=100\mathcal{R}(\delta)$).}\label{fig:largefrequency}
		\label{fig:ref_exact}
	\end{figure}

On figure \ref{fig:largefrequency} we show that the low-frequency part of the time domain solution is actually a good approximation of the full solution, as mentioned in remark \ref{rem:fourier}. It is completely non-trivial, as we have no information on the localisation of poles for the resolvent in the frequency domain outside the \emph{low-frequency} range. It seems that there are no more resonances in the high frequency range due to the dispersive nature of the material. This will be investigated in a future work. 

\subsubsection{About the computational cost}

 We note that, because a small number of modes usually suffices to approximate the reference solution, the computation cost of the asymptotic solution is relatively cheap. The time needed to compute the reference solution and the asymptotic one are linear in $L(= 10^4)$ and $J(= 30)$, respectively. Thus, the time to compute the asymptotic solution is much smaller than the time to compute the reference solution, namely, hundred time smaller. Moreover,  the modes can be pre-calculated and one can compute for a very low cost the response of the particle to any given illumination in the time domain.

\section{Concluding remarks}
In this paper, we have shown that it is possible to define quasi-normal modes (similar to the ones found in the physics literature) for \emph{small} plasmonic particles using the spectral decomposition of the Neumann-Poincar\'e operator and some perturbative spectral analysis. We have proved that, in a three-dimensional setting, only a few modes are necessary to represent the solutions of the scattering problem by a strictly convex plasmonic particle and that these types of representations can give a very good approximation of the field in the time domain. Our numerical simulations have corroborated the validity of this approach in the two-dimensional case. This theoretical and numerical framework can be adapted to handle more complex systems with multiple particles (see \cite{ammari2017mathematicalscalar}).
This work needs to be extended to solutions of Maxwell's equations and to dielectric structures. This will be the subject of forthcoming papers.

\section*{Acknowledgement}
This work was supported by the Swiss National Science Foundation grant number 200021-172483. The authors thank Habib Ammari for helpful conversations.

\section*{Data availability}
The data supporting the findings of this study were generated through \textsc{matlab} and are available from the corresponding author on request. 

	\pagebreak
	\appendix
	
	\section{Properties of the layer potentials}
	We briefly recall here some basic properties of layer potential. There is an abundant literature on the subject. For more details we refer to the books \cite{ammari2013mathematical,nedelec2001acoustic,colton2013integral,kang07}.
	
	\subsection{Definitions and notations}
	\label{subsec:def}
	\begin{definition}\label{de:greenfuction} 
		Denote by $\Gamma^k$ the outgoing Green's function for the homogeneous medium, i.e., the unique solution of the Helmholtz operator:
		\begin{equation*}
		\left(\Delta +k^2\right)\Gamma^k(\cdot,y)=\delta_y(\cdot) \quad \tin \R^d
		\end{equation*} satisfying the Sommerfeld radiation condition. In three dimensions, $\Gamma^k$ is given by
		\begin{equation*}
		\Gamma^k(x,y)=-\frac{e^{ik|x-y|}}{4\pi|x-y|}, \qquad x,y \in \R^3.
		\end{equation*}
		In two dimensions, it is given by
		\begin{equation*}
		\Gamma^{k}(x,y)=
		\begin{dcases}
		\frac{1}{2\pi}\log|x-y|, &\mbox{if }k=0, \\
		-\frac{i}{4}H_0^{(1)}{(k|x-y|)}, &\mbox{if }k>0,
		\end{dcases}
		\end{equation*}
		for $x,y \in \R^2$, where $H_0^{(1)}$ is the well-known Hankel function of the first kind and order $0$.
	\end{definition}
	
		\begin{lemma} \label{lem:Hess} The Hessian matrix of the outgoing fundamental solution in three dimensions
		$\Dbf_x^2 \Gamma^k (x,z)=(D)_{p,q=1}^3$ is with entries
		\begin{align*}
		&D_{pp}=\frac{e^{ik|x-z|}}{4\pi|x-z|^5}\left[|x-z|^2-3(x_p-z_p)^2+3ik(x_p-z_p)^2|x-z|+k^2 |x-z|^2-ik |x-z|^3\right], \\
		&D_{pq}=\frac{e^{ik|x-z|}}{4\pi|x-z|^5}(x_p-z_p)(x_q-z_q)\left[-3+3ik|x-z|+k^2|x-z|^2\right], \qquad \text{for~} p\neq q.
		\end{align*}
	\end{lemma}
	
	\begin{definition}\label{de:layerpotential}
		For a function $\phi \in L^2(\partial D)$, we define the single-layer potential by
		\begin{equation*}
		\mathcal{S}^{k}_D[\phi](x)=\int_{\partial D} \Gamma^k(x,y)\phi(y)\mathrm{d}\sigma(y), \qquad x \in \R^d,
		\end{equation*}
		and the Neumann-Poincar\'e operator by
		\begin{equation*}
		\mathcal{K}^{k,*}_D[\phi](x)=\int_{\partial D} \frac{\partial \Gamma^k(x,y)}{\partial \nu(x)}\phi(y)\mathrm{d}\sigma(y), \qquad x \in \partial D,
		\end{equation*}
	\end{definition}
	When $k=0$, we just write $\mathcal{S}_D$ and $\mathcal{K}_D^*$ for simplicity.

	\subsection{The Calder\'on identity and symmetrisation of $\mathcal{K}_D^*$}
	\label{subsec:symm}
	\begin{lemma}\label{lem:symmetrisation} We recall the following classical results \cite{khavinson2007poincare,ando2016analysis,nedelec2001acoustic}.
		\begin{enumerate}
			\item The following Plemelj's symmetrisation principle identity (also
			known as Calder\'on) holds: 
			\begin{equation} \label{eq:calderon}
			\mathcal{K}_D \mathcal{S}_D = \mathcal{S}_D \mathcal{K}_D^* \qquad \mbox{on~} H^{-\frac{1}{2}}(\partial D).
			\end{equation} 
			\item If $\partial D\in C^{1,\alpha}$, for some $\alpha>0$, then $\mathcal{K}_D^*$ is compact. Let $(\lambda_j,\varphi_j)_{j\in \N}$, be the eigenvalues and normalised eigenfunctions of $\mathcal{K}_D^*$ in $\mathcal{H}^*(\partial D)$. Then $\lambda_j \in ]-1/2, 1/2]$, $\lambda_0=1/2$ and $\lambda_j \rightarrow 0$ as $j\rightarrow \infty$.
			\item The operator $\mathcal{K}_D^*$ is self-adjoint in the Hilbert space $\mathcal{H}^*(\partial D)$ which is $H^{-\frac{1}{2}}(\partial D)$ equipped with the following inner product:
			\begin{align*}
			\left\langle u, v\right\rangle_{\mathcal{H}^*(\partial D)} = 
			\begin{cases}
			\displaystyle -\langle u, \widetilde{\mathcal{S}}_D[v]\rangle_{-\frac{1}{2},\frac{1}{2}}, & d=2, \\
			\displaystyle -\left\langle u, \mathcal{S}_D[v]\right\rangle_{-\frac{1}{2},\frac{1}{2}}, & d=3,
			\end{cases}
			\end{align*}
			where
			\[
			\widetilde{\mathcal{S}}_D[v] = 
			\begin{cases}
			\mathcal{S}_D[v] & \text{if } \langle v, \chi(\partial D)\rangle_{-\frac{1}{2}, \frac{1}{2}} = 0,\\
			-\chi(\partial D) & \text{if } v = \varphi_0,
			\end{cases}
			\]
			with $\varphi_0$ being the unique (in the case of a single particle) eigenfunction of $\mathcal{K}^*_D$ associated with eigenvalue $1/2$ such that $\langle \varphi_0, \chi(\partial D)\rangle_{-\frac{1}{2}, \frac{1}{2}}=1$. Also, $\left\langle \cdot,\cdot\right\rangle_{-\frac{1}{2},\frac{1}{2}}$ is the duality pairing between $H^{-\frac{1}{2}}(\partial D)$ and $H^{\frac{1}{2}}(\partial D)$.
			\item From \cite{ando2016analysis}, we have the following extension of \eqref{eq:calderon} in two dimensions:
			\begin{equation*}
			\mathcal{K}_D \widetilde{\mathcal{S}}_D = \widetilde{\mathcal{S}}_D \mathcal{K}_D^*, \qquad \mbox{on~} H^{-\frac{1}{2}}(\partial D).
			\end{equation*} 
			\item Since $\mathcal{K}_D\left[\chi\left(\partial D\right)\right]=\dfrac{1}{2} \chi(\partial D)$, it holds that
			\begin{equation*}
			\int_{\partial D} \varphi_j =0, \qquad  \mbox{for } j\neq 0.
			\end{equation*}
			\item The following trace formulae hold for $\phi \in H^{-\frac{1}{2}}(\partial D)$:
			\begin{eqnarray*}
				\mathcal{S}^k_D[\phi]|_+&=&\mathcal{S}^k_D[\phi]|_- ,\\
				\left.\frac{\partial \mathcal{S}^k_D[\phi]}{\partial \nu}\right|_{\pm}&=&\left(\pm\frac{1}{2}I+\mathcal{K}^{k,*}_D\right)[\phi],
			\end{eqnarray*}
			where $I$ denotes the identity operator.
			\item The following representation formula holds:
			\begin{align*}
			\mathcal{K}_D^*[\phi]=\sum_{j=0}^\infty \lambda_j \left\langle \phi,\varphi_j\right\rangle_{\mathcal{H}^*(\partial D)} \varphi_j, \qquad \forall\phi \in \mathcal{H}^*(\partial D).
			\end{align*}
		\end{enumerate}
	\end{lemma}
	
	\subsection{Invertibility of the boundary operators}
	\begin{lemma}\label{lem:invertibility3D}
		For $k$ small enough, the three-dimensional single-layer potential $\mathcal{S}^k_D:H^{-1/2}(\partial D)\rightarrow H^{1/2}(\partial D)$ is invertible. $\mathcal{S}_D$ is also invertible.
	\end{lemma}
		\begin{lemma}\label{lem:invertibility3D_0}
$\mathcal{S}_D:H^{-1/2}(\partial D)\rightarrow H^{1/2}(\partial D)$ is invertible in three dimensions.
	\end{lemma}
	In two dimensions, the single-layer potential $\mathcal{S}_D:H^{-1/2}(\partial D)\rightarrow H^{1/2}(\partial D)$ is, in general, not invertible. All the proofs for the following lemmas can be found in \cite{ammari2017mathematicalscalar}.
	\begin{lemma}
		For $k$ small enough, the two-dimensional boundary operator $\widehat{\mathcal{S}}_D^k:\mathcal{H}^*(\partial D)\rightarrow\mathcal{H}^*(\partial D)$ defined as
		\begin{equation}
		\widehat{\mathcal{S}}_D^{k}[\phi](x)=\mathcal{S}_D^0[\phi](x)+\eta_k\int_{\partial D}\phi(y)\mathrm{d}\sigma(y),
		\end{equation}
		is invertible and
		\begin{equation}
		\left(\widehat{\mathcal{S}}_D^k\right)^{-1}=\widetilde{\mathcal{S}}_D^{-1}-\left\langle\widetilde{\mathcal{S}}_D^{-1}[\cdot],\varphi_0\right\rangle_{\mathcal{H}^*(\partial D) }\varphi_0-\mathcal{U}_k,
		\end{equation}
		where $\mathcal{U}_k=\dfrac{\left\langle\widetilde{\mathcal{S}}_D^{-1}[\cdot],\varphi_0\right\rangle_{\mathcal{H}^*(\partial D)}}{\mathcal{S}_D[\varphi_0]+\eta_k} \varphi_0$ and $\eta_k=(1/2\pi)(\log{k}+\gamma-\log{2})-i/4,$
		with the constant $\gamma$ being the Euler constant. Note that $\mathcal{U}_k=\mathcal{O}(1/ \log k)$.
	\end{lemma}
	
	\begin{lemma}\label{lem:invertibility2D}
		For $k$ small enough, the two-dimensional single-layer potential $\mathcal{S}_D^k:\mathcal{H}^*(\partial D)\rightarrow\mathcal{H}^*(\partial D)$ is invertible.
	\end{lemma}

	\section{Scaling properties for a finite volume particle}
	\label{sec:scaling}
	For each function $f$ defined on $\partial D$, we define a corresponding function on $\partial B$ by $\widetilde{f}(X)=f(z+\delta X)$.
	\begin{lemma}
		\label{lem:scaling_1}
		It holds that
		\begin{eqnarray*}
			\mathcal{K}^{k,*}_D[f](x)&=&\mathcal{K}^{k\delta,*}_B[\widetilde{f}](X),\\
			\mathcal{S}^{k}_D[f](x)&=&\delta\mathcal{S}^{k\delta}_B[\widetilde{f}](X).
		\end{eqnarray*}
	\end{lemma}
	
	\begin{lemma}
		\label{lem:scaling_2}
		For $f,g$ defined on $\partial D$, corresponding to $\widetilde{f},\widetilde{g}$, respectively, we have
		\begin{eqnarray*}
			\left\langle f,g \right\rangle_{\mathcal{H}^*(\partial D)}&=& \begin{cases} 
				\delta^2\left\langle \widetilde{f},\widetilde{g} \right\rangle_{\mathcal{H}^*(\partial B)}, & d=2,\\
				\delta^3\left\langle \widetilde{f},\widetilde{g} \right\rangle_{\mathcal{H}^*(\partial B)}, & d=3,
			\end{cases}  
			\\
			||f||_{\mathcal{H}^*(\partial D)}&=&
			\begin{cases} 
				\delta||\widetilde{f}||_{\mathcal{H}^*(\partial B)}, & d=2, \\	
				\delta^{3/2}||\widetilde{f}||_{\mathcal{H}^*(\partial B)}, & d=3.
			\end{cases} 
		\end{eqnarray*}
	\end{lemma}
	\begin{proof}
		In three dimensions, by straightforward calculations we have
		\[
		\begin{aligned} 
		\left\langle f,g \right\rangle_{\mathcal{H}^*(\partial D)} & = \int_{\partial D} f(x) \int_{\partial D} \frac{g(y)}{4\pi|x-y|} \mathrm{d}\sigma(y) \mathrm{d}\sigma(x) \\
		& = \delta^3 \int_{\partial B} f(z+\delta X) \int_{\partial B} \frac{g(z+\delta Y)}{4 \pi|X-Y|} \mathrm{d}\sigma(Y) \mathrm{d}\sigma(X) \\
		& = \delta^3 \left\langle \widetilde{f},\widetilde{g} \right\rangle_{\mathcal{H}^*(\partial B)}.
		\end{aligned} 
		\] 
		Hence, $||f||_{\mathcal{H}^*(\partial D)} = \delta^{3/2}||\widetilde{f}||_{\mathcal{H}^*(\partial B)}$.
		
		In the two-dimensional case we write $\mathcal{H}^*(\partial D) = \mathcal{H}_0^*(\partial D) \oplus \{\mu\varphi_0, \; \mu \in \mathbb{C}\}$ and treat both cases: $g$ belongs to either $\mathcal{H}_0^*(\partial D)$ or $\{\mu\varphi_0, \; \mu \in \mathbb{C}\}$. In the former case, we have
		\[
		\begin{aligned} 
		\left\langle f,g \right\rangle_{\mathcal{H}^*(\partial D)} & = -\frac{1}{2\pi}\int_{\partial D} f(x) \int_{\partial D} g(y) \log(|x-y|) \mathrm{d}\sigma(y) \mathrm{d}\sigma(x) \\
		& = -\frac{\delta^2}{2\pi} \int_{\partial B} f(z+\delta X) \int_{\partial B} g(z+\delta Y) (\log(\delta) + \log(|X-Y|)) \mathrm{d}\sigma(Y) \mathrm{d}\sigma(X) \\
		& = \delta^2 \left\langle \widetilde{f},\widetilde{g} \right\rangle_{\mathcal{H}^*(\partial B)}.
		\end{aligned} 
		\]  
		If $g = \mu\varphi_0$, we have  
		\[
		\begin{aligned} 
		\left\langle f,g \right\rangle_{\mathcal{H}^*(\partial D)} & = \int_{\partial D} \mu f(x) \mathrm{d}\sigma(x) \\
		& = \delta \int_{\partial B}  \mu f(z+\delta X) \mathrm{d}\sigma(X) \\
		& = \delta^2 \left\langle \widetilde{f},\widetilde{g} \right\rangle_{\mathcal{H}^*(\partial B)},
		\end{aligned} 
		\]  
		where the last equality follows from the fact that $\delta \widetilde{\varphi}_0$ is the (unique) eigenfunction of $\mathcal{K}^*_B$ associated with eigenvalue $1/2$ such that $\langle \delta \widetilde{\varphi}_0, \chi(\partial B) \rangle_{-\frac{1}{2}, \frac{1}{2}}=1$. Hence, $||f||_{\mathcal{H}^*(\partial D)} = \delta||\widetilde{f}||_{\mathcal{H}^*(\partial B)}$.
		
	\end{proof}

	\section{Asymptotic expansions}
	
	\subsection{Asymptotic expansions of the boundary operators}
	\label{subsec:exp_operators}
	
	\begin{lemma}\label{lem:exp_op}
	\begin{enumerate}
	\item The three-dimensional single-layer potential and its inverse admit the following expansions in the quasi-static limit $k\delta \rightarrow 0$:
			\begin{eqnarray*}
				\mathcal{S}^{k\delta}_B & = & \mathcal{S}_B+k\delta \mathcal{S}_{B,1}+\left(k\delta\right)^2\mathcal{S}_{B,2}+\mathcal{O}\left(\left(k\delta\right)^3\right), \\
				\left(\mathcal{S}^{k\delta}_B\right)^{-1} & = & \mathcal{S}^{-1}_B+k\delta\mathcal{B}_{B,1}+\left(k\delta\right)^2\mathcal{B}_{B,2}+\mathcal{O}\left(\left(k\delta\right)^3\right),
			\end{eqnarray*}
			where, for $\phi \in H^{-\frac{1}{2}}(\partial B)$,
			\begin{equation*}
			\mathcal{S}_{B,j}[\phi](x)=-\frac{i}{4\pi}\int_{\partial B} \frac{(i|x-y|)^{j-1}}{j!}\phi(y)\mathrm{d}\sigma(y), \qquad x \in \R^3,
			\end{equation*}
			for $j\in \N$. Also $\mathcal{B}_{B,1}=-\mathcal{S}^{-1}_B\mathcal{S}_{B,1}\mathcal{S}^{-1}_B$ and $\mathcal{B}_{B,2}=-\mathcal{S}^{-1}_B\mathcal{S}_{B,2}\mathcal{S}^{-1}_B+\mathcal{S}^{-1}_B\mathcal{S}_{B,1}\mathcal{S}^{-1}_B\mathcal{S}_{B,1}\mathcal{S}^{-1}_B$.
			
			\item The two-dimensional single-layer potential and its inverse admit the following expansions in the quasi-static limit $k\delta \rightarrow 0$:
			\begin{eqnarray*}
				\mathcal{S}^{k\delta}_B&=&\widehat{\mathcal{S}}_B^{k\delta}+(k\delta )^2\log (k\delta) \mathcal{S}_{B,1}^{(1)} +\mathcal{O}\left(\left(k\delta\right)^2\right), \\
				\left(\mathcal{S}^{k\delta}_B\right)^{-1}&=& \mathcal{L}_B+\mathcal{U}_{k\delta} - (k\delta)^2\log (k\delta) \mathcal{L}_B\mathcal{S}_{B,1}^{(1)} \mathcal{L}_B +\mathcal{O}\left(\left(k\delta\right)^2\right),
			\end{eqnarray*}
			where, for $\phi \in H^{-\frac{1}{2}}(\partial B)$,
			\begin{eqnarray*}	
				\mathcal{S}_{B,1}^{(1)}[\phi](x)&=&-\frac{1}{8\pi} \int_{\partial B}|x-y|^{2}\phi(y)\mathrm{d}\sigma(y), \qquad x \in \R^2.\\
			\end{eqnarray*}
			Also $\mathcal{P}_{\mathcal{H}_0^*}$ is the orthogonal projection onto $\mathcal{H}_0^*$, $\mathcal{L}_B = \mathcal{P}_{\mathcal{H}_0^*}\widetilde{\mathcal{S}}^{-1}_B$.
			\item The Neumann-Poincar\'e operator in three dimensions admits the following expansion in the quasi-static limit
			\begin{equation*}
			\mathcal{K}^{k\delta,*}_B=\mathcal{K}^*_B+\left(k\delta\right)^2\mathcal{K}^{*}_{B,2}+\mathcal{O}\left(\left(k\delta\right)^3\right),
			\end{equation*}
			where, for $\phi \in H^{-\frac{1}{2}}(\partial B)$,
			\begin{equation*}
			\mathcal{K}^{*}_{B,2}[\phi](x)=\frac{1}{8\pi }\int_{\partial B}\frac{(x-y)\cdot\nu(x)}{|x-y|}\phi(y)\mathrm{d}\sigma(y), \qquad x \in \partial B.
			\end{equation*}
			\item The Neumann-Poincar\'e operator in two dimensions admits the following expansion in the quasi-static limit
			\begin{equation*}
			\mathcal{K}^{k\delta,*}_B=\mathcal{K}^*_B+ (k\delta)^2 \log (k\delta) \mathcal{K}^{(1)}_{B,1} +\mathcal{O}\left(\left(k\delta\right)^2\right),
			\end{equation*}
			where, for $\phi \in H^{-\frac{1}{2}}(\partial B)$,
			\begin{equation*}
			\mathcal{K}^{(1)}_{B,1}[\phi](x)=-\frac{1}{8\pi}  \int_{\partial B} \frac{\partial |x-y|^{2}}{\partial \nu_x}\phi(y)\mathrm{d}\sigma(y), \qquad x \in \partial B.
			\end{equation*}
		\end{enumerate}
	\end{lemma}
	\begin{proof}
		The proof can be found in \cite{ammari2017mathematicalscalar}.
		
	\end{proof}

	\subsection{Proof of lemma~\ref{le:small_vol_A}}
	\label{subsec:A}
	\begin{proof}
		Recall that
		\begin{equation*}
		\mathcal{A}_B^{\omega\delta/c}=\frac{1}{\varepsilon_m}\left(\frac{1}{2}I+\mathcal{K}^{k_m\delta,*}_B\right)+\frac{1}{\varepsilon_c}\left(\frac{1}{2}I-\mathcal{K}^{k_c\delta,*}_B\right)\left(\mathcal{S}^{k_c\delta}_B\right)^{-1}\mathcal{S}^{k_m\delta}_B. 
		\end{equation*}
		For the three-dimensional case, using lemma~\ref{lem:exp_op} we have by straightforward calculations
		\begin{eqnarray*}
			\mathcal{A}_B^{\omega\delta/c}&=&\frac{1}{\varepsilon_m}\left[\frac{1}{2}I+\mathcal{K}_B^*+(k_m\delta)^2\mathcal{K}_{B,2}\right]+\frac{1}{\varepsilon_c}\left[\frac{1}{2}I-\mathcal{K}_B^*-(k_c\delta)^2\mathcal{K}_{B,2}\right]\left[\mathcal{S}_B^{-1}+(k_c\delta)\mathcal{B}_{B,1}+(k_c\delta)^2\mathcal{B}_{B,2}\right],\\
			&~&\left[\mathcal{S}_B+(k_m\delta)\mathcal{S}_{B,1}+(k_m\delta)^2\mathcal{S}_{B,2}\right]+\mathcal{O}\left(\left(\omega\delta c^{-1}\right)^3\right), \\
			&=&\frac{1}{2}\left(\frac{1}{\varepsilon_c}+\frac{1}{\varepsilon_m}\right)I-\left(\frac{1}{\varepsilon_c}-\frac{1}{\varepsilon_m}\right)\mathcal{K}^*_B+\left(\frac{(k_m\delta)^2}{\varepsilon_m}-\frac{(k_c\delta)^2}{\varepsilon_c}\right)\mathcal{K}_{B,2}+\frac{1}{\varepsilon_c}\left(\frac{1}{2}-\mathcal{K}_B^*\right)\bigg((k_m\delta)\mathcal{S}_B^{-1}\mathcal{S}_{B,1},\\
			&~& +(k_m\delta)^2\mathcal{S}_B^{-1}\mathcal{S}_{B,2}-k_c\delta\mathcal{S}_B^{-1}\mathcal{S}_{B,1}-k_c k_m\delta^2\mathcal{S}_B^{-1}\mathcal{S}_{B,1}\mathcal{S}_B^{-1}\mathcal{S}_{B,1}+(k_c\delta)^2\mathcal{S}_B^{-1}\mathcal{S}_{B,1}\mathcal{S}_B^{-1}\mathcal{S}_{B,1},\\
			&~& -(k_c\delta)^2\mathcal{S}_B^{-1}\mathcal{S}_{B,2}\bigg)+\mathcal{O}\left(\left(\omega\delta c^{-1}\right)^3\right),\\
			&=&\mathcal{A}_B^0+\frac{1}{\varepsilon_c}\left(\frac{1}{2}I-\mathcal{K}_B^*\right)(k_m^2-k_c^2)\delta^2\mathcal{S}_B^{-1}\mathcal{S}_{B,2}
			+\mathcal{O}\left(\left(\omega\delta c^{-1}\right)^3\right),\\
			&=&\mathcal{A}_B^0+\left(\omega \delta c^{-1}\right)^2\frac{\varepsilon_m-\varepsilon_c}{\varepsilon_m\varepsilon_c}\left(\frac{1}{2}I-\mathcal{K}_B^*\right)\mathcal{S}_B^{-1}\mathcal{S}_{B,2}
			+\mathcal{O}\left(\left(\omega\delta c^{-1}\right)^3\right),
		\end{eqnarray*}
		where we used $\mathcal{S}_B^{-1}\mathcal{S}_B=I$ and $$\left(\frac{1}{2}I-\mathcal{K}_B^*\right)\mathcal{S}_B^{-1}\mathcal{S}_{B,1}=0.$$
		For the two-dimensional case, we have
\begin{eqnarray*}
\frac{1}{2}I - \mathcal{K}^{k_c \delta,*}_B &=& \frac{1}{2}I - \mathcal{K}_B^* - (k_c \delta)^2 \log (k_c \delta) \mathcal{K}_{B,1}^{(1)} + \mathcal{O}\left((k_c\delta)^2\right),\\
\left(\mathcal{S}^{k_c \delta}_B\right)^{-1} &=& \mathcal{L}_B + \mathcal{U}_{k_c \delta} - \left(\omega \delta c^{-1}\right)^2 \log \left(\omega \delta c^{-1}\right) \frac{\varepsilon_c}{\varepsilon_m} \mathcal{L}_B \mathcal{S}^{(1)}_{B,1} \mathcal{L}_B + \mathcal{O}\left(\left(\omega \delta c^{-1}\right)^2\right),\\
\mathcal{S}^{k_m \delta}_B &=& \widetilde{S}_B + \Upsilon_{k_m\delta} +\left(\omega \delta c^{-1}\right)^2 \log \left(\omega \delta c^{-1}\right) \mathcal{S}^{(1)}_{B,1} + \mathcal{O}\left(\left(\omega \delta c^{-1}\right)^2\right).
\end{eqnarray*}
		Also, $\mathcal{L}_B \Upsilon_{k_m\delta} = \mathcal{P}_{\mathcal{H}^*_0}\widetilde{\mathcal{S}}_B^{-1} \Upsilon_{k_m\delta} =0$, where 
		\[
		\Upsilon_{k_m\delta}[\psi] = \langle \psi, \widetilde{\phi}_0 \rangle_{\mathcal{H}^*} (\mathcal{S}_B[\widetilde{\phi}_0] +\chi(\partial B) + \eta_{k_m \delta}).
		\]
		Hence,
		\begin{eqnarray*}
			(\mathcal{S}_B^{k_c\delta})^{-1}\mathcal{S}_B^{k_m\delta}=\mathcal{P}_{\mathcal{H}^*_0}+\mathcal{U}_{k_c\delta} \widetilde{\mathcal{S}}_B + \mathcal{U}_{k_c\delta} \Upsilon_{k_m \delta} +  \left(\omega \delta c^{-1}\right)^2 \log  \left(\omega \delta c^{-1}\right) \mathcal{L}_B \mathcal{S}^{(1)}_{D,1} \left(I - \frac{\varepsilon_c}{\varepsilon_m} \mathcal{P}_{\mathcal{H}^*_0}\right) + \mathcal{O}\left(\left(\omega \delta c^{-1}\right)^2\right).
		\end{eqnarray*}
		We have that  
		\[
		\left(\frac{1}{2}I-\mathcal{K}_B^*\right)\mathcal{U}_{k_c\delta} = \dfrac{\left\langle\widetilde{\mathcal{S}}_B^{-1}[\cdot],\widetilde{\phi}_0\right\rangle_{\mathcal{H}^*(\partial B)}}{\mathcal{S}_B[\widetilde{\phi}_0]+\eta_{k_c}} \left(\frac{1}{2}\widetilde{\phi}_0 - \mathcal{K}_B^*[\widetilde{\phi}_0]\right) = 0. 
		\]
	\end{proof}

	\subsection{Proof of lemma~\ref{lem:lemma_F}}
	\label{subsec:exp_f}
	\begin{proof}
		Using the Taylor expansion $$\left.\Gamma^{k_m}(z+\delta X,y)\right|_{X\in\partial B}=\Gamma^{k_m}(z,y)+\delta X\cdot\nabla\Gamma^{k_m}(z,y)+\delta^2 X^\top \Dbf_X^2 \Gamma^{k_m} (z,y)X+... ,$$ we compute, for $X\in \partial B$,
		\begin{eqnarray*}
			\widetilde{F}(X)&=&\widetilde{F}_2(X)+\frac{1}{\delta \varepsilon_c}\left(\frac{1}{2}I-\mathcal{K}^{k_c\delta,*}_B\right)\left(\mathcal{S}^{k_c\delta}_B\right)^{-1}[\widetilde{F}_1](X) \\
			&=&-\frac{1}{\delta\varepsilon_m}\frac{\partial \Gamma^{k_m}(z+\delta X,y)}{\partial \nu_X}-\frac{1}{\delta\varepsilon_c}\left(\frac{1}{2}I-\mathcal{K}^*_B+\mathcal{O}\left((k_c\delta)^2\right)\right)\left(\mathcal{S}^{-1}_B+k_c\delta\mathcal{B}_{B,1}+\mathcal{O}\left((k_c\delta)^2\right)\right)\\
			&~&\left[\Gamma^{k_m}\left(z+\delta X,y\right)\right]\\
			&=&-\frac{1}{\delta\varepsilon_m}\nu_X\cdot\nabla\Gamma^{k_m}(z+\delta X,y)-\frac{1}{\delta\varepsilon_c}\left(\frac{1}{2}I-\mathcal{K}^*_B\right)\mathcal{S}^{-1}_B\left[\Gamma^{k_m}(z+\delta X,y)\right]\\
			&~&-\frac{k_c}{\varepsilon_c}\left(\frac{1}{2}I-\mathcal{K}^*_B\right)\mathcal{B}_{B,1}\left[\Gamma^{k_m}(z+\delta X,y)\right]+\mathcal{O}\left(\omega^2 \delta c ^{-2}\right)\\
			&=&- \frac{1}{\varepsilon_m}\nu_X\cdot\nabla\Gamma^{k_m}(z,y)-\frac{\Gamma^{k_m}(z,y)}{\delta\varepsilon_c}\left(\frac{1}{2}I-\mathcal{K}^*_B\right)\mathcal{S}^{-1}_B\left[\chi(\partial B)\right]\\
			&~&-\frac{\nabla\Gamma^{k_m}(z,y)}{\varepsilon_c}\cdot\left(\frac{1}{2}I-\mathcal{K}^*_B\right)\mathcal{S}^{-1}_B[X]+\mathcal{O}\left(\omega^2 \delta c ^{-2}\right)\\
			&=&-\frac{1}{\varepsilon_m}\nu_X\cdot\nabla\Gamma^{k_m}(z,y)+\frac{1}{\varepsilon_c}\nu_X\cdot\nabla\Gamma^{k_m}(z,y)+\mathcal{O}\left(\omega^2 \delta c ^{-2}\right),
		\end{eqnarray*}
		where we used $\left(\frac{1}{2}I-\mathcal{K}^*_B\right)\mathcal{S}^{-1}_B[\chi(\partial B)]=0$ and $ \left(\frac{1}{2}I-\mathcal{K}^*_B\right)\mathcal{B}_{B,1}=0$. It is immediate to see that $\left(\frac{1}{2}I-\mathcal{K}^*_B\right)\mathcal{S}^{-1}_B[X]=-\nu_X$, indeed assuming there exists $\phi \in \mathcal{H}^*(\partial B)$ such that $\mathcal{S}^{-1}_B[X]=\phi$, then $\mathcal{S}_B[\phi]=X$, and $\left.\partial \mathcal{S}_B[\phi]/\partial \nu_X \right |_-= \nu_X$ which is equivalent to $\left(\frac{1}{2}I-\mathcal{K}^*_B\right)[\phi]=-\nu_X$ using jump conditions.
	\end{proof}
	
	\subsection{Proof of lemma~\ref{lem:lemma_F_2D}}
\label{subsec:exp_f_2D}
\begin{proof}
Using the Taylor expansion
$$
e^{ik_m d\cdot (z + \delta X)} = e^{ik_m d\cdot z} +  \frac{i \omega}{c} [d \cdot \delta X]e^{ik_m d\cdot z} + \mathcal{O}\left(\left(k_m\delta\right)^2\right),
$$
we compute, for $X \in \partial B$,
\begin{eqnarray*}
	\widetilde{F}(X)&=&\widetilde{F}_2(X)+\frac{1}{\delta \varepsilon_c}\left(\frac{1}{2}I-\mathcal{K}^{k_c\delta,*}_B\right)\left(\mathcal{S}^{k_c\delta}_B\right)^{-1}[\widetilde{F}_1](X) \\
	&=&-\frac{1}{\delta\varepsilon_m}\frac{\partial e^{ik_m d\cdot (z + \delta X)}}{\partial \nu_X}-\frac{1}{\delta\varepsilon_c}\left(\frac{1}{2}I-\mathcal{K}^*_B + \mathcal{O}\left(\left(k_c \delta\right)^2 \log(k_c \delta)\right)\right)\left(\mathcal{L}_B+\mathcal{U}_{k_c\delta}+ \mathcal{O}\left(\left(k_c \delta\right)^2 \log(k_c \delta)\right)\right)\\
	&~&\left[e^{ik_m d\cdot (z + \delta X)}\right]\\
	&=&-\frac{1}{\delta\varepsilon_m}\nu_X\cdot\nabla e^{ik_m d\cdot (z + \delta X)}-\frac{1}{\delta\varepsilon_c}\left(\frac{1}{2}I-\mathcal{K}^*_B\right)\widetilde{\mathcal{S}}^{-1}_B\left[ e^{ik_m d\cdot (z + \delta X)}\right] +\mathcal{O}\left(\omega^2 \delta c ^{-2} \log \left(\omega \delta c^{-1}\right)\right)\\
	&=&- \frac{1}{\varepsilon_m}\nu_X\cdot\nabla e^{ik_m d\cdot z}-\frac{e^{ik_m d\cdot z}}{\delta\varepsilon_c}\left(\frac{1}{2}I-\mathcal{K}^*_B\right)\widetilde{\mathcal{S}}^{-1}_B\left[\chi(\partial B)\right]\\
	&~&-\frac{\nabla e^{ik_m d\cdot z}}{\varepsilon_c}\cdot\left(\frac{1}{2}I-\mathcal{K}^*_B\right)\widetilde{\mathcal{S}}^{-1}_B[X]+\mathcal{O}\left(\omega^2 \delta c ^{-2} \log \left(\omega \delta c^{-1}\right)\right)\\
	&=&-\frac{1}{\varepsilon_m}\nu_X\cdot\nabla e^{ik_m d\cdot z}+\frac{1}{\varepsilon_c}\nu_X\cdot\nabla e^{ik_m d\cdot z}+\mathcal{O}\left(\omega^2 \delta c ^{-2} \log \left(\omega \delta c^{-1}\right)\right),
\end{eqnarray*}
		where we used $\left(\frac{1}{2}I-\mathcal{K}^*_B\right)\widetilde{\mathcal{S}}^{-1}_B[\chi(\partial B)]=0$ and $ \left(\frac{1}{2}I-\mathcal{K}^*_B\right)\mathcal{U}_{k \delta}=0$. It is immediate to see that $\left(\frac{1}{2}I-\mathcal{K}^*_B\right)\widetilde{\mathcal{S}}^{-1}_B[X] =-\nu_X$, indeed assuming there exists $\phi \in \mathcal{H}^*(\partial B)$ such that $\widetilde{\mathcal{S}}^{-1}_B[X]=\phi$, then $\widetilde{\mathcal{S}}_B[\phi]=X$ so $\left.\partial \widetilde{\mathcal{S}}_B[\phi]/\partial \nu_X \right |_-= \nu_X$ and using  jump conditions for $\phi \neq \varphi_0$ we get $\left.\partial \widetilde{\mathcal{S}}_B[\phi]/\partial \nu_X \right |_-=\left.\partial \mathcal{S}_B[\phi]/\partial \nu_X \right |_-=\left(-\frac{1}{2}I+\mathcal{K}^*_B\right)[\phi]=\nu_X$.

\end{proof}

\section{Proof of theorem~\ref{theo:resonanceexpansion2}}\label{sec:appendixproof2D}
We need the following lemma:
\begin{lemma}
	The Hankel function has the following asymptotics as $x\rightarrow +\infty$:
	\begin{equation}
	H_0^{(1)}(x)=\sqrt{\frac{2}{\pi x}}e^{i(x-\pi/4)}+\mathcal{O}\left(\frac{1}{x}\right).
	\end{equation}
\end{lemma}

For $x$ large and $y \in \partial D$:
\begin{align*}
\mathcal{S}_D^{\frac{\omega}{c}}\left[\varphi_j\right](x) =& - \frac{i}{4} \int_{\partial D} H_0^{(1)}\left(\omega c^{-1}|x-y|\right)\varphi_j (y) \dd \sigma (y) \\
\sim &- \frac{i\sqrt{2}}{4\sqrt{\pi}}   \int_{\partial D} \frac{e^{i\left(\omega c^{-1} |x-y|-\pi/4\right)}}{\sqrt{\omega c^{-1} |x-y|}}\varphi_j (y) \dd \sigma (y).
\end{align*} 

\begin{lemma}\label{lem:lemma_F_2D} 
As $\omega\delta c^{-1} \rightarrow 0$, $F$ defined in \eqref{eq:F_scalar} admits the following asymptotic expansion: 
	\begin{equation*}
	F(x) = \frac{f(\omega)}{\delta} \left[i \omega\delta c^{-1} e^{i \omega c^{-1} d\cdot z}\left(\frac{1}{\varepsilon_c}-\frac{1}{\varepsilon_m}\right) d \cdot \nu_x +\mathcal{O}\left(\left(\omega\delta c^{-1} \right)^2 \log \left(\omega\delta c^{-1} \right)\right)\right], \quad x \in \partial D.
	\end{equation*}
\end{lemma}
\begin{proof}
See appendix~\ref{subsec:exp_f_2D}.
\end{proof}

\begin{lemma}\label{lem:u_sca_2D} 
As $\omega\delta c^{-1} \rightarrow 0$, the scalar field admits the following asymptotic expansion: 
	\begin{equation*}
u^\text{sca}(x,\omega) \approx \frac{e^{-i\pi/4}}{\sqrt{8\pi c}} \sum_{j=1}^J \left\langle d \cdot \nu,\varphi_j\right\rangle_{\mathcal{H}^*(\partial D)} \Xi_j(x,\omega),
	\end{equation*}
	where the modes $\Xi_j$ are defined by
	\begin{equation*}
	\Xi_j(x,\omega):=\int_{\partial D} \frac{\varphi_j(y)}{\sqrt{|x-y|}} \frac{f(\omega)\sqrt{\omega}}{(\lambda(\omega)-\lambda_j(\omega \delta))} e^{i\omega c^{-1}(|x-y|+d\cdot z)}  \dd \sigma(y),
	\end{equation*}
	and $\lambda_j(\omega \delta) :=  \lambda_j - \left(\omega\delta c^{-1}\right)^2\log{\left(\omega\delta c^{-1}\right)}\alpha_j +\mathcal{O}\left(\left(\omega\delta c^{-1}\right)^2\right)$. 
\end{lemma}

\begin{proof}
Since $\left\langle \nu,\varphi_0\right\rangle_{\mathcal{H}^*(\partial D)}=0$, the zeroth term vanishes in the summation. 
\end{proof}

The goal of this section is to establish a resonance expansion for the low-frequency part of the scattered field in the time domain.
Introduce, for $0<\epsilon<\rho$,  the truncated inverse Fourier transform of the scattered field $u^{\text{sca}}$ given by 
\begin{equation*}
P_{\rho,\epsilon}\left[\widehat{u}^\text{sca}\right](x,t)=\int_{-\rho}^{-\epsilon} u^{\text{sca}}(x,\omega) e^{-i\omega t} \dd\omega+\int_{\epsilon}^{\rho}  u^{\text{sca}}(x,\omega) e^{-i\omega t} \dd\omega.
\end{equation*}

Recall that $z$ is the centre of the resonator and $\delta$ its radius. Let us define $$t_0^\pm(d,x):=\frac{1}{c}\left(|x-z|+d\cdot z \pm 2\delta\right)\pm C_1,$$
the time it takes to the signal to reach first the scatterer and then observation point $x$. The term $\pm 2 \delta/c$ accounts for the maximal timespan spent inside the particle.

\label{app:resonanceexpansion2}
\begin{proof}
We have 
\begin{equation} 
P_{\rho,\epsilon}\left[\widehat{u}^\text{sca}\right](x,t) \sim \sum_{j=1}^J \frac{ e^{-i \pi/4}}{\sqrt{8 \pi c}} \left\langle d \cdot \nu,\varphi_j\right\rangle_{\mathcal{H}^*(\partial D)}\left[\int_{-\rho}^{-\epsilon} \Xi_j(x,\omega)e^{-i\omega t}\dd \omega+\int_{\epsilon}^{\rho} \Xi_j(x,\omega)e^{-i\omega t}\dd \omega\right].
\label{eq:RHS1}
\end{equation} 
	For $j\geq 1$, let us compute the contribution of one mode $\Xi_j(x,\omega)$. We want to apply the residue theorem to get an asymptotic expansion in the time domain. Note that:
	\begin{align*}
	\int_{-\rho}^{-\epsilon} \Xi_j(x,\omega) e^{-i\omega t} \mathrm{d}\omega+\int_{\epsilon}^{\rho} \Xi_j(x,\omega) e^{-i\omega t} \mathrm{d}\omega&=\\
	\oint_{\mathcal{C}^{\pm}} \Xi_j(x,\Omega) &e^{-i\Omega t}\mathrm{d}\Omega -\int_{\mathcal{C}_\rho^{\pm}} \Xi_j(x,\Omega) e^{-i\Omega t}  \mathrm{d}\Omega
	-\int_{\mathcal{C}_\epsilon^{\pm}} \Xi_j(x,\Omega) e^{-i\Omega t}  \mathrm{d}\Omega,
	\end{align*}
	where the integration contours $\mathcal{C}_\rho^{\pm}$ and $\mathcal{C}_\epsilon^{\pm}$ are semi-circular arcs of radius $\rho$ and $\epsilon$, respectively, in the upper (+) or lower (-) half-planes, and $\mathcal{C}^{\pm}$ is the closed contour defined as $\mathcal{C}^{\pm}:=\mathcal{C}_\rho^{\pm}\cup\mathcal{C}_\epsilon^{\pm}\cup[-\rho,-\epsilon]\cup[\epsilon,\rho]$.
	The integral on the closed contour is the main contribution to the scattered field by the mode and can be computed using the residue theorem to get, for $\rho\geq \max_{j \in N}\Re[\Omega_j^\pm(\delta)]$ and $0<\epsilon\leq \min_{j\in N} \Re[\Omega_j^\pm(\delta)]$,
	\begin{align*}
	\oint_{\mathcal{C}^{+}}  \Xi_j(x,\Omega) e^{-i\Omega t} \mathrm{d}\Omega&=0,\\
	\oint_{\mathcal{C}^{-}}  \Xi_j(x,\Omega) e^{-i\Omega t} \mathrm{d}\Omega&=2\pi i\text{Res}\left( \Xi_j(x,\Omega)e^{-i\Omega t},\Omega_j^\pm(\delta)\right).
	\end{align*}
	Since $\Omega_j^\pm(\delta)$ is a simple pole of $\omega \mapsto \dfrac{1}{\lambda(\omega)-\lambda_j(\omega \delta)}$ we can write:
	\begin{align*}
	\oint_{\mathcal{C}^{-}} \Xi_j(x,\Omega) e^{-i\Omega t} \mathrm{d}\Omega&=2\pi i\text{Res}\left(\Xi_j(x,\Omega),\Omega_j^\pm(\delta)\right)e^{-i\Omega_j^\pm(\delta)t}.
	\end{align*}
	To compute the integrals on the semi-circle, we introduce:
	\begin{equation*}
	X_j(y,\Omega)=\frac{\varphi_j(y)}{\sqrt{|x-y|}} \frac{\sqrt{\omega}}{(\lambda(\omega)-\lambda_j)} \qquad y \in \partial D.
	\end{equation*}
	
	%Note that $\Bbf_n(\cdot,\cdot,\Omega)$ behaves like a polynomial in $\Omega$ when $\vert \Omega\vert \rightarrow \infty$.
	Given the regularity of the input signal $\widehat{f} \in C_0^{\infty}([0,C_1])$, the Paley-Wiener theorem~\cite[p.161]{Yosida1995FA} ensures decay properties of its Fourier transform at infinity. For all $N\in\mathbb{N}^*$ there exists a positive constant $C_N$ such that for all $\Omega \in \mathbb{C}$
	\begin{equation*}
	|f(\Omega)|\leq C_N (1+|\Omega|)^{-N}e^{C_1 |\Im{(\Omega)}|}.
	\end{equation*}
		Let $T:=(|x-y|+d\cdot z)/c$. We now rewrite the integrals on the large semi-circle
	\begin{align*}
	\int_{\mathcal{C}_\rho^{\pm}} \Xi_j(x,\Omega) e^{-i\Omega t} \mathrm{d}\Omega=\int_{\mathcal{C}_\rho^{\pm}}f(\Omega) \int_{\partial D} X_j(y,\Omega) e^{i \Omega \left(T-t\right)}\dd \sigma(y) \dd \Omega.
	\end{align*}
	We have that $t_0^-+C_1 \leq T \leq t_0^+-C_1 $.
	Two cases arise. 
	\paragraph{Case 1:}
	For $0<t<t_0^-$ , i.e., when the signal emitted at $s$ has not reached the observation point $x$, we choose the upper-half integration contour $\mathcal{C}^+$. Transforming into polar coordinates, $\Omega=\rho e^{i\theta}$ for $\theta \in [0,\pi]$, we get:
	\begin{align*}
	\left\vert  e^{i \Omega \left(T -t\right)}\right\vert \leq  e^{ -(t_0^--t+C_1)\Im(\Omega)} \qquad \forall y \in \partial D,
	\end{align*}
	and
	\begin{align*}
	\left|\int_{\mathcal{C}_\rho^{+}}  \Xi_j(x,\Omega) e^{-i\Omega t} \mathrm{d}\Omega\right| & \leq \int_0^\pi \rho \left|f\left(\rho e^{i\theta}\right)\right|e^{-\rho (t_0^--t+C_1)\sin{\theta}}\int_{\partial D}\vert X_j(y, \rho e^{i\theta}) \vert \dd \sigma(y) \mathrm{d}\theta,\\
	&\leq \rho C_N(1+\rho)^{-N} \delta\max_{\theta\in [0,\pi]}{\left \vert X_j\left(\cdot, \rho e^{i\theta}\right) \right\vert_{L^{\infty}(\partial D)}} \pi \frac{1-e^{-\rho(t^-_0-t)}}{\rho(t^-_0-t)},
	\end{align*}
	where we used that for $\theta \in [0,\pi/2]$, we have $\sin{\theta} \geq 2\theta/\pi \geq 0$ and $-\cos{\theta}\leq-1+2\theta/\pi$. The usual way to go forward from here is to take the limit $\rho \rightarrow \infty$, and get that the limit of the integral on the semi-circle is zero. As in the three-dimensional case, we work in the quasi-static approximation here, and our modal expansion is not uniformly valid for all frequencies. So we have to work with a fixed maximum frequency $\rho$. However, the maximum frequency $\rho$ depends on the size of the particle via the hypothesis $\rho \leq c\delta^{-1}$. Since $N$ can be taken arbitrarily large and that $X_j$ behaves like a polynomial in $\rho$ \emph{whose degree does not depend on $j$}, we get that, uniformly for $j\in [1,J]$:
	\begin{align*}
	\left|\int_{\mathcal{C}_\rho^{+}} \Xi_j(x,\Omega) e^{-i\Omega t} \mathrm{d}\Omega\right| = \mathcal{O}\left(\frac{\delta}{t_0^--t} \rho^{-N}\right).
	\end{align*}
	
		For the upper-half semi-circle of radius $\epsilon$, we also transform into polar coordinates with the change of variable $\Omega=\epsilon e^{i\theta}$, for $\theta \in [0,\pi]$, and get:
	\begin{equation*}
	\left|\int_{\mathcal{C}_\epsilon^{+}}  \Xi_j(x,\Omega) e^{-i\Omega t} \mathrm{d}\Omega\right|  \leq \epsilon C_N(1+\rho)^{-N} \delta\max_{\theta\in [0,\pi]}{\left \vert X_j\left(\cdot, \epsilon e^{i\theta}\right) \right\vert_{L^{\infty}(\partial D)}} \pi \frac{1-e^{-\epsilon [(t^-_0-t)]}}{\epsilon (t^-_0-t)},
	\end{equation*}
	
	\paragraph{Case 2:}
	For $t>t_0^+$ , we choose the lower-half integration contour $\mathcal{C}^-$. Transforming into polar coordinates, $\Omega=\rho e^{i\theta}$ for $\theta \in [\pi,2\pi]$, we get
	\begin{align*}
	\left\vert  e^{i \Omega \left(T -t\right)}\right\vert \leq  e^{ (t-t_0^+-C_1) \Im (\Omega)} \qquad \forall y \in \partial D^2,
	\end{align*}
	and
	\begin{align*}
	\left|\int_{\mathcal{C}_\rho^{-}} \Xi_j(x,\Omega) e^{-i\Omega t} \mathrm{d}\Omega\right| & \leq \int_\pi^{2\pi}\rho \left|f\left(\rho e^{i\theta}\right)\right|e^{\rho (t-t_0^+)\sin{\theta}}\int_{\partial D}\vert X_j(y, \rho e^{i\theta}) \vert \dd \sigma(y) \dd \theta,\\
	&\leq \rho C_N(1+\rho)^{-N}\delta \max_{\theta\in [\pi,2\pi]}{\left\vert X_j\left(\cdot, \rho e^{i\theta}\right) \right\vert_{L^{\infty}( \partial D)}}\pi \frac{1-e^{-\rho(t-t_0^+)}}{\rho(t-t_0^+)},
	\end{align*}
	Exactly as in Case $1$, we cannot take the limit $\rho \rightarrow \infty$. However, the maximum frequency $\rho$ depends on the size of the particle via the hypothesis $\rho \leq c\delta^{-1} $. Using the fact that $N$ can be taken arbitrarily large and that $X_j$ behaves like a polynomial in $\rho$ \emph{whose degree does not depend on $j$}, we get that, uniformly for $j\in [1,J]$: 
	\begin{align*}
	\left|\int_{\mathcal{C}_\rho^{-}} \Xi_j(x,\Omega) e^{-i\Omega t} \mathrm{d}\Omega\right| = \mathcal{O}\left(\frac{\delta}{t} \rho^{-N}\right).
	\end{align*}
	
	For the lower-half semi-circle of radius $\epsilon$, we also transform into polar coordinates with the change of variable $\Omega=\epsilon e^{i\theta}$, for $\theta \in [0,\pi]$, and get:
	\begin{equation*}
	\left|\int_{\mathcal{C}_\epsilon^{-}}  \Xi_j(x,\Omega) e^{-i\Omega t} \mathrm{d}\Omega\right|  \leq \epsilon C_N(1+\rho)^{-N} \delta\max_{\theta\in [\pi,2\pi]}{\left \vert X_j\left(\cdot, \epsilon e^{i\theta}\right) \right\vert_{L^{\infty}(\partial D)}} \pi \frac{1-e^{-\epsilon (t-t^+_0)}}{\epsilon (t-t^+_0)},
	\end{equation*}	
	The result of theorem~\ref{theo:resonanceexpansion2} is obtained by summing the contribution of all the modes. 
\end{proof}

	\bibliographystyle{siam}
	\bibliography{references}
\end{document}